\documentclass[11pt]{article}

\usepackage{epic}
\usepackage{eepic}
\usepackage{latexsym}
\usepackage{epsf}

\newtheorem{theorem}{Theorem}[section]
\newtheorem{lemma}[theorem]{Lemma}
\newtheorem{claim}[theorem]{Claim}

\newtheorem{corol}[theorem]{Corollary}
\newtheorem{definition}[theorem]{Definition}

\newenvironment{proof}{{\bf Proof:\ }}{\hfill$\Box$\medskip}

\newcommand{\ignore}[1]{}

\ignore{
\newcommand{\RANDSHORTPATH}
{\mbox{\bf RAND-}\allowbreak\mbox{\bf SHORT-}\allowbreak\mbox{\bf PATH}}
\newcommand{\SHORTPATH}{\mbox{\bf SHORT-PATH}}
\newcommand{\UNWEIGHTEDSHORTPATH}{\mbox{\bf UNWGHT-SHORT-PATH}}
\newcommand{\APPROXSHORTPATH}
{\mbox{\bf APPROX-}\allowbreak{\bf SHORT-}\allowbreak\mbox{\bf PATH}}
\newcommand{\APPROXDISTPROD}
{\mbox{\bf APPROX-}\allowbreak\mbox{\bf DIST-}\allowbreak\mbox{\bf PROD}}
\newcommand{\DISTPROD}{\mbox{\bf DIST-PROD}}
\newcommand{\FASTPROD}{\mbox{\bf FAST-PROD}}
\newcommand{\SCALE}{\mbox{\bf SCALE}}
\newcommand{\RAND}{\mbox{\bf RAND}}
\newcommand{\FINDBRIDGE}{\mbox{\bf FIND-BRIDGE}}
\newcommand{\FINDBRIDGEUPD}{\mbox{\bf FIND-BRIDGE-UPD}}
\newcommand{\PATH}{\mbox{\bf PATH}}
\newcommand{\SPATH}{\mbox{\bf S-PATH}}
\newcommand{\SUBPATH}{\mbox{\bf SUB-PATH}}
\newcommand{\SUBPATHUPD}{\mbox{\bf SUB-PATH-UPD}}
\newcommand{\CC}{{\cal C}}
\newcommand{\HITTINGSET}{\mbox{\bf HITTING-SET}}

\newcommand{\WITTOSUC}{\mbox{\bf WIT-to-SUC}}
\newcommand{\DISTPRODUPD}{\mbox{\bf DIST-PROD-UPD}}
\newcommand{\UPDATE}{\mbox{\bf UPDATE}}
}

\newcommand{\RANDSHORTPATH}
{\mbox{\bf rand-}\allowbreak\mbox{\bf short-}\allowbreak\mbox{\bf path}}
\newcommand{\SHORTPATH}{\mbox{\bf short-path}}
\newcommand{\UNWEIGHTEDSHORTPATH}{\mbox{\bf unwght-short-path}}
\newcommand{\APPROXSHORTPATH}
{\mbox{\bf approx-}\allowbreak{\bf short-}\allowbreak\mbox{\bf path}}
\newcommand{\APPROXDISTPROD}
{\mbox{\bf approx-}\allowbreak\mbox{\bf dist-}\allowbreak\mbox{\bf prod}}
\newcommand{\DISTPROD}{\mbox{\bf dist-prod}}
\newcommand{\FASTPROD}{\mbox{\bf fast-prod}}
\newcommand{\SCALE}{\mbox{\bf scale}}
\newcommand{\RAND}{\mbox{\bf rand}}
\newcommand{\FINDBRIDGE}{\mbox{\bf find-bridge}}
\newcommand{\FINDBRIDGEUPD}{\mbox{\bf find-bridge-upd}}
\newcommand{\PATH}{\mbox{\bf path}}
\newcommand{\SPATH}{\mbox{\bf s-path}}
\newcommand{\SUBPATH}{\mbox{\bf sub-path}}
\newcommand{\SUBPATHUPD}{\mbox{\bf sub-path-upd}}
\newcommand{\CC}{{\cal C}}
\newcommand{\HITTINGSET}{\mbox{\bf hitting-set}}

\newcommand{\WITTOSUC}{\mbox{\bf wit-to-suc}}
\newcommand{\DISTPRODUPD}{\mbox{\bf dist-prod-upd}}
\newcommand{\UPDATE}{\mbox{\bf update}}

\newcommand{\algorithm}{{\tt algorithm}}

\newcommand{\ct}{\!\cdot\!}
\newcommand{\Ot}{\tilde{O}}
\newcommand{\del}{\delta}

\newcommand{\eps}{\epsilon}
\newcommand{\bit}{{\sl bit}}
\newcommand{\eq}{\;=\;}
\begin{document}


\title{
All Pairs Shortest Paths using Bridging Sets \\
and Rectangular Matrix Multiplication 
 \thanks{A preliminary version of
 this paper appeared in \protect\cite{Zw98FO}.}
}

\author{
{\em Uri Zwick}
 \thanks{Department of Computer Science,  Tel Aviv University, Tel Aviv
 69978, Israel. E-mail address: {\tt zwick@math.tau.ac.il}.
 Work supported in part by THE ISRAEL SCIENCE FOUNDATION founded by
 The Israel Academy of Sciences and Humanities. }
}

\maketitle


\begin{abstract}
{
\noindent
\setlength{\parindent}{0pt}%
\setlength{\parskip}{5pt plus 1pt}%
We present two new algorithms for solving the {\em All
Pairs Shortest Paths\/} (APSP) problem for weighted directed
graphs. Both algorithms use fast matrix multiplication algorithms.

\noindent
The first algorithm 
solves the APSP problem for weighted directed graphs in which the edge
weights are integers of small absolute value in $\Ot(n^{2+\mu})$ time,
where $\mu$ satisfies the equation $\omega(1,\mu,1)=1+2\mu$ and
$\omega(1,\mu,1)$ is the exponent of the multiplication of an $n\times
n^\mu$ matrix by an $n^\mu \times n$ matrix. Currently, the best
available bounds on $\omega(1,\mu,1)$, obtained by Coppersmith,
imply that $\mu<0.575$. The running time of our algorithm is therefore
$O(n^{2.575})$. Our algorithm improves on the $\Ot(n^{(3+\omega)/2})$
time algorithm, where $\omega=\omega(1,1,1)<2.376$ is the usual exponent
of matrix multiplication, obtained by Alon, Galil and Margalit, whose
running time is only known to be $O(n^{2.688})$.

The second algorithm 
solves the APSP problem {\em almost\/} exactly for directed graphs with
{\em arbitrary\/} non-negative real weights. The algorithm runs in
$\Ot((n^\omega/\eps)\log (W/\eps))$ time, where $\eps>0$ is an error
parameter and~$W$ is the largest edge weight in the graph, after the
edge weights are scaled so that the smallest non-zero edge
weight in the graph is~1. It returns estimates of all the distances in
the graph with a stretch of at most $1+\eps$. Corresponding paths can
also be found efficiently.
}
\end{abstract}



\section{Introduction}

The {\em All Pairs Shortest Paths\/} (APSP) problem is one of the most
fundamental algorithmic graph problems. The complexity of the fastest
known algorithm for solving the problem for weighted directed graphs
with arbitrary real weights is $O(mn+n^2\log n)$, where~$n$ and~$m$,
respectively, are the number of vertices and edges in the graph.  This
algorithm is composed of a preliminary step, due to Johnson \cite{Jo77},
in which cycles of negative weight are found and eliminated, and a
nonnegative weight function that induces the same shortest paths is
found. The algorithm then proceeds by running Dijkstra's single source
shortest paths algorithm (Dijkstra \cite{Di59}), implemented using
Fibonacci heaps (Fredman and Tarjan \cite{FrTa87}), from each vertex of
the graph. For a clear description of the whole algorithm see Cormen,
Leiserson and Rivest \cite{CoLeRi90}, Chapters~21, 25 and~26.

For directed graphs with nonnegative edge weights, the running time of
the above algorithm can be reduced to $O(m^*n+n^2\log n)$, where~$m^*$
is the number of edges that participate in shortest paths (Karger,
Koller and Phillips \cite{KaKoPh93}, and McGeoch \cite{McGeoch95}).  For
undirected graphs with nonnegative integer edge weights, a running time
of $O(mn)$ can be obtained by running a recent single source shortest
paths algorithm of Thorup \cite{Thorup99},\cite{Thorup00} from each
vertex of the graph.

\ignore{
If the edge weights are positive or negative polynomially-sized
integers, then scaling techniques of Gabow and Tarjan \cite{GaTa89}
and Goldberg \cite{Goldberg95}, combined with the above mentioned
algorithms, yield an $O(m^*n+n^2\log n + m\sqrt{n})$ algorithm for the
problem.
}

The running time of all the above mentioned algorithms may be as high as
$\Omega(n^3)$. Can the APSP problem be solved in sub-cubic time? Fredman
\cite{Fredman76} showed that the APSP problem for weighted directed graphs
can be solved {\em non-uniformly\/} in $O(n^{2.5})$ time. 
More precisely, for every~$n$, there is a program that solves the APSP
problem for graphs with~$n$ vertices using at most $O(n^{2.5})$ comparisons,
additions and subtractions. But, a separate program has to be used for 
each input size. Furthermore, the size of the program that works on graphs 
with~$n$ vertices may be exponential in~$n$. Fredman used this result to
obtain a uniform algorithm that runs in $O(n^3((\log\log n)/\log
n)^{1/3})$ time.
Takaoka \cite{Takaoka92} slightly improved this bound to
$O(n^3((\log\log n)/\log n)^{1/2})$. These running times are just barely
sub-cubic.

The APSP problem is closely related to the problem of computing the
min/plus product, or {\em distance product\/}, as we shall call it, of
two matrices. If $A=(a_{ij})$ and $B=(b_{ij})$ are two $n\times n$
matrices, then their distance product $C=A\star B$ is an $n\times n$
matrix $C=(c_{ij})$ such that $c_{ij} = \min_{k=1}^n\, \{
a_{ik}+b_{kj}\}$, for $1\le i,j\le n$. A weighted graph $G=(V,E)$ on~$n$
vertices can be encoded as an $n\times n$ matrix $D=(d_{ij})$ in
which~$d_{ij}$ is the weight of the edge $(i,j)$, if there is such an
edge in the graph, and $d_{ij}=+\infty$, otherwise. We also let
$d_{ii}=0$, for $1\le i\le n$. It is easy to see that~$D^n$, the $n$-th
power of~$D$ with respect to distance products, is a matrix that
contains the distances between all pairs of vertices in the graph
(assuming there are no negative cycles). The matrix~$D^n$ can be
computed using $\lceil\log_2 n\rceil$ distance products. It is, in fact,
possible to show that the distance matrix~$D^n$ can be computed in
essentially the same time required for just one distance product (see
\cite{AHU74}, Section~5.9).

Two $n\times n$ matrices over a {\em ring\/} can be multiplied using
$O(n^\omega)$ algebraic operations, where $\omega$ is the exponent of
square matrix multiplication. The naive matrix multiplication algorithm
shows that $\omega\le 3$. The best upper bound on $\omega$ is currently
$\omega<2.376$ (Coppersmith and Winograd \cite{CopWin90}). The only
lower bound available on $\omega$ is the naive lower bound $\omega\ge
2$.  Unfortunately, the fast matrix multiplication
algorithms cannot be used directly to compute distance products,
as the set of integers, or the set of reals, is not a ring with respect to
the operations $\min$ and plus.

Alon, Galil and Margalit \cite{AlGaMa97} were the first to show that
fast matrix multiplications algorithms can be used to obtain improved
algorithms for the APSP problem for graphs with small integer edge
weights. They obtained an algorithm whose running time is
$\Ot(n^{(3+\omega)/2})$ \footnote{Throughout the paper, $\Ot(f(n))$
   stands for $O(f(n)\log^c n)$, for some $c>0$.} for solving the APSP
problem for directed graphs with edge weights taken from the set
$\{-1,0,1\}$. Galil and Margalit \cite{GaMa97a},\cite{GaMa97b} and
Seidel \cite{Seidel95}, obtained $\Ot(n^\omega)$ time algorithms for
solving the APSP problem for unweighted {\em undirected\/} graphs.
Seidel's algorithm is much simpler. The algorithm of Galil and Margalit
has the advantage that it can be extended to handle small integer
weights. The running time of their algorithm, when used to solve the
APSP problem for undirected graphs with edge weights taken from the set
$\{0,1,\ldots,M\}$, is $\Ot(M^{(\omega+1)/2}n^\omega)$. An improved time
bound of $\Ot(Mn^\omega)$ for the same problem was recently obtained by
Shoshan and Zwick \cite{ShZw99}.

In this paper we present an improved algorithm for solving the APSP
problem for directed graphs with edge weights of small absolute value.
The improved efficiency is gained by using {\em bridging sets\/} and by
using {\em rectangular\/} matrix multiplications instead of square
matrix multiplications, as used by Alon, Galil and Margalit
\cite{AlGaMa97}. It is possible to reduce a rectangular matrix
multiplication into a number of square matrix multiplications. For
example, the task of computing the product of an $n\times m$ matrix by
an $m \times n$ matrix is easily reduced to the task of computing
$(n/m)^2$ products of two $m\times m$ matrices.  The running time of our
algorithm, if we use this approach, is $\Ot(n^{2+{1}/{(4-\omega)}})$,
which is $O(n^{2.616})$, if we use the estimate $\omega<2.376$.
However, a more efficient implementation is obtained if we compute the
rectangular matrix multiplications directly using the fastest
rectangular matrix multiplication algorithms available.  The running
time of the algorithm is then $\Ot(n^{2+\mu})$, where $\mu$ satisfies
the equation $\omega(1,\mu,1)=1+2\mu$, where $\omega(1,\mu,1)$ is the
exponent of the multiplication of an $n\times n^\mu$ matrix by an $n^\mu
\times n$ matrix. \footnote{In general, $\omega(r,s,t)$ is the exponent
   of the multiplication of an $n^s\times n^r$ matrix by an $n^r\times
   n^t$ matrix.} Currently, the best available bounds on
$\omega(1,\mu,1)$, obtained by Coppersmith \cite{Coppersmith97} and by
Huang and Pan \cite{HuPa98}, imply that $\mu<0.575$. The running time of
our algorithm is therefore $O(n^{2.575})$, and possibly better.

If $\omega=2$, as may turn out to be the case, then the running time of
both our algorithm and the algorithm of Alon, Galil and Margalit, would
be $\Ot(n^{2.5})$. However, the running time of our algorithm may be
$\Ot(n^{2.5})$ even if $\omega>2$. To show that the running time of our
algorithm is $\Ot(n^{2.5})$ it is enough to show that
$\omega(1,\frac{1}{2},1)=2$, i.e., that the product of an $n\times
n^{1/2}$ matrix by an $n^{1/2}\times n$ matrix can be performed in
$\Ot(n^2)$ time. Coppersmith \cite{Coppersmith97} showed that the
product of an $n\times n^{0.294}$ by an $n^{0.294}\times n$ matrix can
be computed in $\Ot(n^2)$ time.

The algorithm of Alon, Galil and Margalit \cite{AlGaMa97} can also handle
integer weights taken from the set $\{-M,\ldots,0,\ldots,M\}$, i.e.,
integer weights of absolute value at most~$M$. The running time of their
algorithm is then
$\Ot(M^{{(\omega-1)}/2}n^{(3+\omega)/2})$, if $M\le
n^{(3-\omega)/(\omega+1)}$, and  $\Ot(Mn^{{(5\omega-3)}/{(\omega+1)}})$,
if $n^{(3-\omega)/(\omega+1)}\le M$. Takaoka \cite{Takaoka98} obtained an
algorithm whose running time is $\Ot(M^{1/3} n^{(6+\omega)/3})$. The
bound of Takaoka is better than the bound of Alon, Galil and Margalit for
larger values of~$M$. The running time of Takaoka's algorithm is sub-cubic
for $M< n^{3-\omega}$.

Our algorithm can also handle small integer weights, i.e.,
weights taken from the set $\{-M,\ldots,0,\ldots,M\}$. If
rectangular matrix multiplications are reduced to square
matrix multiplications, then the running time of the
algorithm is $\Ot(M^{1/(4-\omega)}n^{2+1/(4-\omega)})$. This
running time is again sub-cubic for $M< n^{3-\omega}$ but, for
every $1\le M< n^{3-\omega}$ the running time of our
algorithm is faster than both the algorithms of Alon, Galil
and Margalit and of Takaoka. The running time is further
reduced if the rectangular matrix multiplications required
by the algorithm are computed using the best available
algorithm. If $M=n^t$, where $t<3-\omega$, then the running
time of the algorithm is $\Ot(n^{2+\mu(t)})$, where
$\mu=\mu(t)$ satisfies the equation $\omega(1,\mu,1)=1+2\mu-t$.

The new algorithm for solving the APSP problem for graphs
with small integer weights is extremely simple and natural,
despite the somewhat cumbersome bounds on its running time.
We already noted that to compute
all the distances in a weighed graph on~$n$ vertices represented by the
matrix~$D$ it is enough to square the matrix~$D$ about $\log_2
n$ times with respect to distance products. It turns out
that if we are willing to repeat this process, say, $\log_{3/2}
n$ times, then in the $i$-th iteration, instead of squaring
the current matrix, it is enough to choose a set~$B_i$ of
roughly $m_i=(2/3)^i n$ columns of the current matrix and
multiply them by the corresponding~$m_i$ rows of the
matrix. In fact, a randomly chosen set of about~$m_i$
columns would be a good choice with a very high probability!
We have thus replaced the product of two $n\times n$
matrices in the $i$-th iteration by a product of an $n\times
m_i$ matrix by an $m_i\times n$ matrix.

To convert distance products of matrices into normal algebraic products
we use a technique suggested in \cite{AlGaMa97} (see also Takaoka
\cite{Takaoka98}), based on a previous idea of Yuval \cite{Yuval76}.
Suppose that $A=(a_{ij})$ and $B=(b_{ij})$ are two $n\times n$ matrices
with elements taken from the set $\{-M,\ldots,0,\ldots,M\}$. We
convert~$A$ and $B$ into two $n\times n$ matrices $A'=(a'_{ij})$ and
$B'=(b'_{ij})$ where $a'_{ij}=(n+1)^{M-a_{ij}}$ and
$b'_{ij}=(n+1)^{M-b_{ij}}$. It is not difficult to see that the distance
product of~$A$ and $B$ can be inferred from the algebraic product
of~$A'$ and~$B'$ (see the next section). We pay, however, a high price
for this solution. Each element of~$A'$ and~$B'$ is a huge number that
about $M\log n$ bits, or about~$M$ words of $\log n$ bits each, are
needed for its representation. An algebraic operation on elements of the
matrices~$A'$ and~$B'$ {\em cannot\/} be viewed therefore as a single
operation. Each such operation can be carried out, however, in
$\Ot(M\log n)$ time. We would have to take this factor into account in
our complexity estimations.

Our results indicates that it may be possible to solve the APSP problem
for directed graphs with small integer weights
{\em uniformly\/} in $\Ot(n^{2.5})$ time. Even if this were
the case, there would still be a gap between the
complexities of the directed
and undirected versions of the APSP problem. As mentioned,
the APSP for {\em undirected\/} graphs with small integer
weights can be solved in $\Ot(n^\omega)$ time, as shown by
Seidel \cite{Seidel95} and by 
Galil and Margalit \cite{GaMa97a},\cite{GaMa97b}. (See also Shoshan and
Zwick \cite{ShZw99}.)

We next show that the gap between the directed and the
undirected versions of the APSP problem can be closed if we are
willing to settle for {\em approximate\/} shortest paths.
We say that a path between two vertices $i$ and $j$ is of
stretch $1+\epsilon$ if its length is at most $1+\eps$ times
the distance from $i$ to $j$. It is fairly easy to see that
paths of stretch $1+\eps$ between all pairs of vertices of
an {\em unweighted\/} directed graph can be computed in
$\Ot(n^\omega/\eps)$ time. (This fact is mentioned in
\cite{GaMa97a}). 
Stretch 2 paths, or at least
stretch~2 distances, for example, may be
obtained by computing the matrices $A^{2^r}$, for $1\le r\le
\lceil\log_2 n\rceil$, where~$A$ is the adjacency matrix of the graph, and
Boolean products are used this time.

We extend this result and obtain an algorithm for finding stretch
$1+\epsilon$ paths between all pairs of vertices of a directed graph
with {\em arbitrary non-negative\/} real weights. The running time of
the algorithm is $\Ot((n^\omega/\eps)\ct\log (W/\eps))$, where~$W$ is the
largest edge weight in the graph after the edge weights are scaled so
that the smallest non-zero edge weight is~1. Our algorithm uses a simple
{\em adaptive scaling\/} technique. It is observed by Dor, Halperin and
Zwick \cite{DoHaZw00} that for any $c\ge 1$, computing stretch~$c$
distances between all pairs of vertices in an unweighted directed graph
on~$n$ vertices is at least as hard as computing the Boolean product of
two $n/3\times n/3$ matrices.  Our result is therefore very close to
being optimal.

Algorithms for approximating the distances between all pairs
of vertices in a weighted {\em undirected\/} graph were
obtained by Cohen and Zwick \cite{CoZw97}. They present an
$\Ot(n^2)$ algorithm for finding paths with stretch at most 3, an
$\Ot(n^{7/3})$ algorithm for finding paths of stretch $7/3$,
and an $\Ot(n^{3/2}m^{1/2})$ algorithm for finding paths of
stretch $2$. The algorithms of Cohen and Zwick \cite{CoZw97}
use ideas obtained by Aingworth, Chekuri, Indyk and Motwani
\cite{AiChInMo99} and by Dor, Halperin and Zwick
\cite{DoHaZw00} who designed algorithms that approximate
distances in unweighted undirected graphs with a small 
{\em additive\/} error. As can be seen from their running
times, these algorithms are all purely combinatorial. They
do not use fast matrix multiplication algorithms.
It is also observed in \cite{DoHaZw00} that for any
$1\le c<2$, computing stretch~$c$ distances between all pairs of
vertices in an unweighted undirected graph on~$n$ vertices
is again at least as hard as computing the Boolean product
of two $n/3\times n/3$ matrices. For $\eps<1$, our
algorithm is therefore close to optimal even for undirected graphs. 

\ignore{
The algorithms presented in this extended abstract are initially
designed to compute or approximate the {\em distances\/} in the graph.
We then show how to modify them so that they would also return a
representation of shortest paths or small-stretch paths between all
pairs of vertices in the graph.  The main tool used for this purpose is
an algorithm for finding {\em witnesses\/} for Boolean matrix
multiplication. A simple randomized algorithm for finding witnesses is
described by Seidel \cite{Seidel95}.  A deterministic algorithm,
obtained by derandomizing Seidel's algorithm was obtained by Alon and
Naor \cite{AlNa96}. Another deterministic algorithm for finding
witnesses for Boolean matrix multiplication was obtained by Galil and
Margalit \cite{GaMa93}. We need to extend these results and obtain an
algorithm for finding witnesses for distance products.
}

The rest of the paper is organized as follows. In the next section we
present an algorithm that uses fast matrix multiplication to speed up
the computation of distance products. In Section~\ref{S-witness} we
introduce the notion of {\em witnesses\/} for distance products. Such
witnesses are used to reconstruct shortest paths. In
Section~\ref{S-exact} we present a simple {\em randomized\/} algorithm
for solving the APSP problem in directed graphs with small integer
weights. In Section~\ref{S-deter} we introduce the notion of {\em
   bridging sets\/} and explain how the randomized algorithm of the
previous section can be converted into a deterministic algorithm, if the
input graph is unweighted. A deterministic algorithm for weighted graphs
is then given in Section~\ref{S-weighted}. In
Section~\ref{S-approx} we present the new algorithm for obtaining an
almost exact solution to the APSP problem for directed graphs with
arbitrary non-negative real weights.
Finally, we end in Section~\ref{S-concl} with some concluding remarks
and open problems.

\section{Distance product of matrices}
\label{S-prod}

We begin with a definition of distance products.

\begin{definition}[Distance products]
Let~$A$ be an $n\times m$ matrix and $B$ be an $m\times n$ matrix.
The {\em distance product\/}
of~$A$ and $B$, denoted $A\star B$, in an $n\times n$ matrix~$C$ such that
$$c_{ij}\, = \,\min_{k=1}^m\, \{ a_{ik}+b_{kj}\}\;, 
{\rm \ for \ } 1\le i,j\le n\;.$$
\end{definition}

In this definition, and in the rest of
the paper, we use the convention that matrices are denoted by upper case
letters, and that the elements of a matrix are denoted by the
corresponding lower case letter.

The distance product of~$A$ and $B$ can be computed naively in $O(n^2
m)$ time. Alon, Galil and Margalit \cite{AlGaMa97} (see also Takaoka
\cite{Takaoka98}) describe a way of using fast matrix multiplication,
and fast integer multiplication, to compute distance products of
matrices whose elements are taken from the set
$\{-M,\ldots,0,\ldots,M\}\cup\{+\infty\}$ . The running time of their
algorithm, when applied to rectangular matrices, is
$\Ot(Mn^{\omega(1,r,1)})$, where $m=n^r$. Here $O(n^{\omega(1,r,1)})$ is
the number of algebraic operations required to compute the standard
algebraic product of an $n\times n^r$ matrix by an $n^r\times n$ matrix.
We see, therefore, that the running time of this algorithm depends
heavily on~$M$. For large values of~$M$ the naive algorithm, whose
running time is independent of~$M$, is faster.

Algorithm \DISTPROD, whose description is given in Figure~\ref{F-prod},
uses the faster of these two methods to compute the distance product of
an $n\times m$ matrix~$A$ and an $m\times n$ matrix~$B$ whose elements
are integers. We let $m=n^r$. Elements in~$A$ and~$B$ that are of
absolute value greater than~$M$ are treated as if they were~$+\infty$.
(This feature is used by the algorithms described in the subsequent
sections.) Algorithm \FASTPROD, called by \DISTPROD, computes the
algebraic product of two integer matrices using the fastest rectangular
matrix multiplication algorithm available, and using the
Sch\"{o}nhage-Strassen \cite{SchSt71} (see also \cite{AHU74}) algorithm
for integer multiplication. 

\begin{figure}[t]
\begin{center}
\framebox{\hspace{0.6cm}\parbox{4.8in}{
{\tt algorithm} \DISTPROD$(A,B,M)$ $\phantom{2^{2^{2^2}}}$ \\[5pt]
{\tt if} $Mn^{\omega(1,r,1)} \le n^{2+r}$\\
{\tt then}\\
$\null\qquad a'_{ij} \gets \cases{(m+1)^{M-a_{ij}} & if $\;|a_{ij}|\le M$ \cr
           \hfill   0  \hfill        & otherwise\cr}$ \\[3pt]
$\null\qquad b'_{ij} \gets \cases{(m+1)^{M-b_{ij}} & if $\;|b_{ij}|\le M$ \cr
           \hfill   0  \hfill        & otherwise\cr}$ \\[7pt]
$\null\qquad C'\gets\FASTPROD(A',B')$ \\[5pt]
$\null\qquad c_{ij}\gets \cases{
2M-\lfloor\log_{(m+1)} c'_{ij}\rfloor & if $\;c'_{ij}>0$\cr 
+\infty & otherwise\cr} $ \\[5pt]
{\tt else}\\
$\null\qquad c_{ij}\, \gets \,\min_{k=1}^m\, \{ a_{ik}+b_{kj}\}\;,
\ignore{\rm \ for \ } 1\le i,j\le n\;.$ \\[3pt]
{\tt endif} \\[3pt]
{\tt return} $C$ $\phantom{2_{2_{2_2}}}$
}\hspace{0.6cm}}
\end{center}

\caption{\label{F-prod}Computing the distance product of two matrices.}
\end{figure}

\begin{lemma}\label{L-dist-prod}
    Algorithm \DISTPROD\ computes the distance product of an $n\times
    n^r$ matrix by an $n^r\times n$ matrix whose finite entries are all
    of absolute value at most~$M$ in
    $\Ot(\min\{Mn^{\omega(1,r,1)},n^{2+r}\})$ time.
\end{lemma}

\begin{proof}
If $n^{2+r}<Mn^{\omega(1,r,1)}$ then \DISTPROD\ computes the distance
product of~$A$ and~$B$ using the naive algorithm that runs in
$O(n^{2+r})$ time and we are done.

Assume, therefore, that $Mn^{\omega(1,r,1)}\le n^{2+r}$.
To see that the algorithm correctly computes the distance
product of~$A$ and~$B$ in this case, note that 
for every $1\le i,j\le n$ we have
$$c'_{ij}\;=\;\sum_{k=1}^m {(m+1)}^{2M-(a_{ik}+b_{kj})}\;,$$
where indices $k$ for which $a_{ik}=+\infty$ or $b_{kj}=+\infty$ are
excluded from the summation,
and therefore 
$$c_{ij}\;=\; \min_{k=1}^m \,\{a_{ik}+b_{kj}\} \;=\;
2M-\lfloor\log_{(m+1)} c'_{ij}\rfloor\;.$$
We next turn to the
complexity analysis.  If $Mn^{\omega(1,r,1)} \le n^{2+r}$ then
\FASTPROD\ performs $\Ot(n^{\omega(1,r,1)})$ arithmetical operations on
$O(M\log n)$-bit integers.  The Sch\"{o}nhage-Strassen integer
multiplication algorithm multiplies two $k$-bit integers using $O(k\log
k\log\log k)$ bit operations. Letting $k=O(M\log n)$, we get that the
complexity of each arithmetic operation is $\Ot(M\log n)$. Finally, the
logarithms used in the computation of $c_{ij}$ can be easily implemented
using binary search. The complexity of the algorithm in this case is
therefore $\Ot(Mn^{\omega(1,r,1)})$, as required.
\end{proof}

There is, in fact, a slightly more efficient way of implementing
\FASTPROD. Instead of computing the product of~$A'$ and~$B'$ using
multiprecision integers, we can compute the product of~$A'$ and~$B'$
modulo about~$M$ different prime numbers with about $\log n$ bits each
and then reconstruct the result using the Chinese remainder theorem.
This reduces the running time, however, by only a polylogarithmic factor.

What is known about $\omega(1,r,1)$, the exponent of the multiplication
of an $n\times n^r$ matrix by an $n^r\times n$ matrix? Note that
$\omega=\omega(1,1,1)$ is the famous exponent of (square) matrix
multiplication. The best bound on $\omega$ is currently $\omega<2.376$
(Coppersmith and Winograd \cite{CopWin90}). It is easy to see that a
product of an $n\times n^r$ matrix by an $n^r\times n$ matrix can be
broken into $n^{2(1-r)}$ products of $n^r\times n^r$ matrices, and can
therefore by computed in $O(n^{2+r(\omega-2)})$ time. It follows,
therefore, that $\omega(1,r,1)\le 2+r (\omega-2)$. Better bounds are
known, however. Coppersmith \cite{Coppersmith97} showed that the product
of an $n\times n^{0.294}$ matrix by an $n^{0.294}\times n$ matrix can be
computed using $\Ot(n^2)$ arithmetical operations. Let $\alpha = \sup
\{\, 0\le r\le 1 : \omega(1,r,1)=2+o(1) \}$. It follows from
Coppersmith's result that $\alpha\ge 0.294$. Note that if
$\omega=2+o(1)$, then $\alpha=1$.  An improved bound for
$\omega(1,r,1)$, for $\alpha\le r\le 1$ can be obtained by combining the
bounds $\omega(1,1,1)<2.376$ and $\omega(1,\alpha,1)=2+o(1)$. The
following lemma is taken from Huang and Pan \cite{HuPa98}:

\begin{lemma}\label{L-omega}
Let $\omega=\omega(1,1,1)<2.376$ and let $\alpha=\sup
\{\, 0\le r\le 1 : \omega(1,r,1)=2+o(1) \}>0.294$. Then
$$\omega(1,r,1)\le \cases{ \hfill 2+o(1) \hfill & if $\;0\le r\le \alpha$, \cr
2+\frac{\omega-2}{1-\alpha}(r-\alpha)+o(1) & if $\;\alpha\le r\le 1$.\cr}$$
\end{lemma}

Note that the upper bound on $\omega(1,r,1)$ given in
Lemma~\ref{L-omega} is a piecewise linear function.  (See
Figure~\ref{F-omega} in Section~\ref{S-exact}.)  Another well known fact
(see, e.g., Pan \cite{Pan85} or Burgisser, Clausen and Shokrollahi
\cite{BuClSh97}) regarding matrix multiplication, used in later
sections, is the fact that $\omega(r,s,t)$, the exponent of computing
the product of an $n^r\times n^s$ matrix and an $n^s\times n^t$ matrix,
does not change if the order of its arguments is change. In particular:

\begin{lemma}\label{L-11r}
$\omega(1,1,r)=\omega(1,r,1)=\omega(r,1,1)$.
\end{lemma}

In other words, the cost of computing the product of an
$n\times n^r$ matrix by an $n^r\times n$ matrix, and the cost of
computing the product of an $n\times n$ matrix by an $n\times n^r$
matrix are asymptotically the same.

\section{Witnesses for distance products}
\label{S-witness}

Next, we introduce the notion of {\em witnesses\/} for distance products
of matrices. Witnesses for distance products are used to reconstruct
shortest paths.

\begin{definition}[Witnesses]
    Let~$A$ be an $n\times m$ matrix and $B$ be an $m\times n$ matrix. An
    $n\times n$ matrix~$W$ is said to be a matrix of witnesses for the
    distance product $C=A\star B$ if for every $1\le i,j\le n$ we have
    $1\le w_{ij}\le m$ and $c_{ij} = a_{i,w_{ij}} + b_{w_{ij},j}$.
\end{definition}

Using ideas of Seidel \cite{Seidel95}, Galil and Margalit \cite{GaMa93}
and Alon and Naor \cite{AlNa96}, it is easy to extend algorithm
\DISTPROD\ so that it would also return a matrix of witnesses. The
running time of \DISTPROD\ would increase by only a polylogarithmic
factor. The details are sketched below. 

There is a simple,
but expensive, way of computing witnesses for the distance product
$C=A\star B$, where~$A$ is an $n\times m$ matrix, and~$B$ is an $m\times
n$ matrix. Let $A'=(a'_{ij})$ and $B'=(b'_{ij})$ be matrices such
that $a'_{ij} = ma_{ij}+j-1$ and $b'_{ji}=mb_{ji}$, for every $1\le i\le
n$ and $1\le j\le m$. If we compute the distance product $C'=A'\star
B'$, then $\lfloor C'/m\rfloor$ is the distance product of~$A\star B$
and $(C'\bmod m)+1$ is a corresponding matrix of witnesses. Furthermore,
all the witnesses in this matrix are the {\em smallest possible\/}
witnesses. The drawback of this approach is that the entries of~$A$
and~$B$ are multiplied by~$m$ and this may slow down the operation
\DISTPROD\ by a factor of~$m$, which may be a huge factor.

There is, however, a much more efficient way of finding witnesses.  We
show, at first, how to find witnesses for elements that have {\em
   unique\/} witnesses. For $1\le k\le m$ and $1\le \ell\le \lceil\log_2
m\rceil+1$, we let $\bit_\ell(k)$ be the $\ell$-th bit in the binary
representation of~$k$. (For concretness, assume that $\bit_1(k)$ is the
least significat bit in the representation of~$k$.) For $1\le \ell\le
\lceil\log_2 m\rceil+1$, let $I_\ell = \{ 1\le k\le m \mid
\bit_\ell(k)=1\}$. We also need the following definition which is also
used in subsequent sections:

\begin{definition}[Sampling]\label{D-samp}
    Let $A$ be an $n\times m$ matrix, and let $I\subseteq
    \{1,2,\ldots,m\}$.  Then, $A[*,I]$ is defined to be the matrix
    composed of the columns of~$A$ whose indices belong to~$I$.
    Similarly, if $B$ is an $m\times n$ matrix, then $B[I,*]$ is defined
    to be the matrix composed of the rows of~$B$ whose indices belong
    to~$I$.
\end{definition}

To find witnesses for all elements of $A=B\star C$ that have a unique
witness, we compute the $O(\log m)$ distance products $C_\ell =
A[*,I_\ell]\star B[I_\ell,*]$, for $1\le \ell\le \lceil\log_2
m\rceil+1$. Let $C_\ell = (c^{(\ell)}_{ij})$. It is easy to see that
$c^{(\ell)}_{ij}=c_{ij}$, if and only if there is a
witness for $c_{ij}$ whose $\ell$-th bit is 1. If~$c_{ij}$ has a unique
witness $w_{ij}$, then these conditions can be used to identify the
individual bits in the binary representation of~$w_{ij}$, and hence
$w_{ij}$ itself. Note that we do not have to know in advance whether
$c_{ij}$ has a unique witness. We just reconstruct a candidate witness
$w_{ij}$ and then check whether $c_{ij}=a_{i,w_{ij}}+b_{w_{ij},j}$.

What do we do with elements that have more than one witness? We use
sampling. For every $1\le r\le \log m$, we choose $s = c\log n$ random
subsets $R_{r1},\ldots,R_{rs}$ of $\{1,2,\ldots,m\}$ of size $m/2^r$.
For every such random set~$R_{rt}$, where $1\le r\le \log m$ and $1\le
t\le s$, we try to find unique witnesses for the product
$A[*,R_{rt}]\star B[R_{rt},*]$. When such a witness is found, we check
whether it is also a witness for the original distance product $A\star
B$. A simple calculation, identical to a calculation that appears in
Seidel \cite{Seidel95}, shows that if the constant $c$ is taken to be
large enough, then with very high probability, we will find in this way
witnesses for all positions.

The above discussion gives a randomized algorithm for computing a matrix
of witnesses for the distance product $A\star B$. The randomized
algorithm uses $O(\log^3 n)$ ordinary distance products of matrices of
equal or smaller size. The resulting algorithm can be derandomized using
the results of Alon and Naor \cite{AlNa96}. We thus obtain:

\begin{lemma}
    An extended version of algorithm \DISTPROD\ computes the distance
    product of an $n\times n^r$ matrix by an $n^r\times n$ matrix whose
    finite entries are all of absolute value at most~$M$, and a
    corresponding matrix of witnesses, in
    $\Ot(\min\{Mn^{\omega(1,r,1)},n^{2+r}\})$ time.
\end{lemma}

In the sequel, we let $(C,W)\gets \DISTPROD(A,B,M)$ denote an invocation
of the extended version of \DISTPROD\ that returns the product
matrix~$C$ and a matrix of witnesses~$W$.

\section{A randomized algorithm for finding shortest paths}
\label{S-exact}

A simple randomized algorithm, 
\RANDSHORTPATH, for finding
distances, and a representation of shortest paths, between all pairs of
vertices of a directed graph on~$n$ vertices in which all edge weights
are taken from the set $\{-M,\ldots,0,\ldots,M\}$ is given in
Figure~\ref{F-rand}.

\begin{figure}[t]
\begin{center}\hspace*{-5pt}
\framebox{\hspace{0.4cm}\parbox{4.8in}{
{\tt algorithm} \RANDSHORTPATH$(D)$ $\phantom{2^{2^{2^2}}}$ \\[5pt]
$F\gets D\,$ ; $\,W\gets 0$ \\
$M\gets \max\{\;|d_{ij}| : d_{ij}\ne+\infty\}$\\[3pt]
{\tt for} $\ell\gets 1$ {\tt to} $\lceil\log_{3/2}n\rceil$ {\tt do}\\
{\tt begin}\\[3pt]
$\null\quad s\gets (3/2)^\ell$\\
$\null\quad B\gets \RAND(\{1,2,\ldots,n\},(9\ln n)/s)$\\
$\null\quad (F',W')\gets 
\mbox{$\DISTPROD(F[*,B],F[B,*],sM)\hspace*{-15pt}$}$\\[3pt]
$\null\quad${\tt for every} $1\le i,j\le n$ {\tt do}\\
$\null\quad${\tt if} $f'_{ij}<f_{ij}$ {\tt then} $f_{ij}\gets f'_{ij}$
; $w_{ij}\gets b_{w'_{ij}}$ {\tt endif}\\[3pt]
{\tt end}\\[3pt]
{\tt return} $(F,W)$ $\phantom{2_{2_{2_2}}}$
}\hspace{0.4cm}}
\end{center}

\caption{\label{F-rand}A randomized algorithm for finding shortest paths.}
\end{figure}

The input to \RANDSHORTPATH\ is an $n\times n$ matrix~$D$ that contains the
weights (or lengths) of the edges of the input graph. We assume that the
vertex set of the graph is $V=\{1,2,\ldots,n\}$. The element~$d_{ij}$ is
the weight of the directed edge from~$i$ to~$j$ in the
graph, if there is such an edge, or~$+\infty$, otherwise.

\begin{figure}[t]
\begin{center}
\setlength{\unitlength}{0.00047in} 
\begingroup\makeatletter\ifx\SetFigFont\undefined%
\gdef\SetFigFont#1#2#3#4#5{%
  \reset@font\fontsize{#1}{#2pt}%
  \fontfamily{#3}\fontseries{#4}\fontshape{#5}%
  \selectfont}%
\fi\endgroup%
{\renewcommand{\dashlinestretch}{30}
\begin{picture}(6244,5803)(0,-10)
\path(794,2444)(1694,2444)(1694,44)
	(794,44)(794,2444)
\path(3344,5744)(5744,5744)(5744,3344)
	(3344,3344)(3344,5744)
\path(44,5744)(2444,5744)(2444,3344)
	(44,3344)(44,5744)
\path(3344,794)(3344,1694)(5744,1694)
	(5744,794)(3344,794)
\texture{44555555 55aaaaaa aa555555 55aaaaaa aa555555 55aaaaaa aa555555 55aaaaaa 
	aa555555 55aaaaaa aa555555 55aaaaaa aa555555 55aaaaaa aa555555 55aaaaaa 
	aa555555 55aaaaaa aa555555 55aaaaaa aa555555 55aaaaaa aa555555 55aaaaaa 
	aa555555 55aaaaaa aa555555 55aaaaaa aa555555 55aaaaaa aa555555 55aaaaaa }
\shade\path(344,5744)(494,5744)(494,3344)
	(344,3344)(344,5744)
\path(344,5744)(494,5744)(494,3344)
	(344,3344)(344,5744)
\shade\path(644,5744)(794,5744)(794,3344)
	(644,3344)(644,5744)
\path(644,5744)(794,5744)(794,3344)
	(644,3344)(644,5744)
\shade\path(1244,5744)(1394,5744)(1394,3344)
	(1244,3344)(1244,5744)
\path(1244,5744)(1394,5744)(1394,3344)
	(1244,3344)(1244,5744)
\shade\path(1994,5744)(2144,5744)(2144,3344)
	(1994,3344)(1994,5744)
\path(1994,5744)(2144,5744)(2144,3344)
	(1994,3344)(1994,5744)
\shade\path(1844,5744)(1994,5744)(1994,3344)
	(1844,3344)(1844,5744)
\path(1844,5744)(1994,5744)(1994,3344)
	(1844,3344)(1844,5744)
\shade\path(3344,5444)(5744,5444)(5744,5294)
	(3344,5294)(3344,5444)
\path(3344,5444)(5744,5444)(5744,5294)
	(3344,5294)(3344,5444)
\shade\path(3344,5144)(5744,5144)(5744,4994)
	(3344,4994)(3344,5144)
\path(3344,5144)(5744,5144)(5744,4994)
	(3344,4994)(3344,5144)
\shade\path(3344,4544)(5744,4544)(5744,4394)
	(3344,4394)(3344,4544)
\path(3344,4544)(5744,4544)(5744,4394)
	(3344,4394)(3344,4544)
\shade\path(3344,3944)(5744,3944)(5744,3794)
	(3344,3794)(3344,3944)
\path(3344,3944)(5744,3944)(5744,3794)
	(3344,3794)(3344,3944)
\shade\path(3344,3794)(5744,3794)(5744,3644)
	(3344,3644)(3344,3794)
\path(3344,3794)(5744,3794)(5744,3644)
	(3344,3644)(3344,3794)
\shade\path(944,2444)(1094,2444)(1094,44)
	(944,44)(944,2444)
\path(944,2444)(1094,2444)(1094,44)
	(944,44)(944,2444)
\shade\path(794,2444)(944,2444)(944,44)
	(794,44)(794,2444)
\path(794,2444)(944,2444)(944,44)
	(794,44)(794,2444)
\shade\path(1094,2444)(1244,2444)(1244,44)
	(1094,44)(1094,2444)
\path(1094,2444)(1244,2444)(1244,44)
	(1094,44)(1094,2444)
\shade\path(1244,2444)(1394,2444)(1394,44)
	(1244,44)(1244,2444)
\path(1244,2444)(1394,2444)(1394,44)
	(1244,44)(1244,2444)
\shade\path(1394,2444)(1544,2444)(1544,44)
	(1394,44)(1394,2444)
\path(1394,2444)(1544,2444)(1544,44)
	(1394,44)(1394,2444)
\shade\path(1544,2444)(1694,2444)(1694,44)
	(1544,44)(1544,2444)
\path(1544,2444)(1694,2444)(1694,44)
	(1544,44)(1544,2444)
\shade\path(3344,1694)(5744,1694)(5744,1544)
	(3344,1544)(3344,1694)
\path(3344,1694)(5744,1694)(5744,1544)
	(3344,1544)(3344,1694)
\shade\path(3344,1544)(5744,1544)(5744,1394)
	(3344,1394)(3344,1544)
\path(3344,1544)(5744,1544)(5744,1394)
	(3344,1394)(3344,1544)
\shade\path(3344,1394)(5744,1394)(5744,1244)
	(3344,1244)(3344,1394)
\path(3344,1394)(5744,1394)(5744,1244)
	(3344,1244)(3344,1394)
\shade\path(3344,1244)(5744,1244)(5744,1094)
	(3344,1094)(3344,1244)
\path(3344,1244)(5744,1244)(5744,1094)
	(3344,1094)(3344,1244)
\shade\path(3344,1094)(5744,1094)(5744,944)
	(3344,944)(3344,1094)
\path(3344,1094)(5744,1094)(5744,944)
	(3344,944)(3344,1094)
\shade\path(3344,944)(5744,944)(5744,794)
	(3344,794)(3344,944)
\path(3344,944)(5744,944)(5744,794)
	(3344,794)(3344,944)
\put(2894,4394){\makebox(0,0)[b]{\smash{{{\SetFigFont{10}{12.0}{\rmdefault}{\mddefault}{\updefault}$\star$}}}}}
\put(2744,1169){\makebox(0,0)[b]{\smash{{{\SetFigFont{10}{12.0}{\rmdefault}{\mddefault}{\updefault}$\star$}}}}}
\put(494,1169){\makebox(0,0)[b]{\smash{{{\SetFigFont{10}{12.0}{\rmdefault}{\mddefault}{\updefault}$n$}}}}}
\put(1244,2594){\makebox(0,0)[b]{\smash{{{\SetFigFont{10}{12.0}{\rmdefault}{\mddefault}{\updefault}$|B|$}}}}}
\put(4544,1919){\makebox(0,0)[b]{\smash{{{\SetFigFont{10}{12.0}{\rmdefault}{\mddefault}{\updefault}$n$}}}}}
\put(6044,1094){\makebox(0,0)[b]{\smash{{{\SetFigFont{10}{12.0}{\rmdefault}{\mddefault}{\updefault}$|B|$}}}}}
\end{picture}
}
\end{center}
\caption{\label{F-sample}
Replacing the square product $F\star F$ by the rectangular product 
$F[*,B]\star F[B,*]$.}
\end{figure}
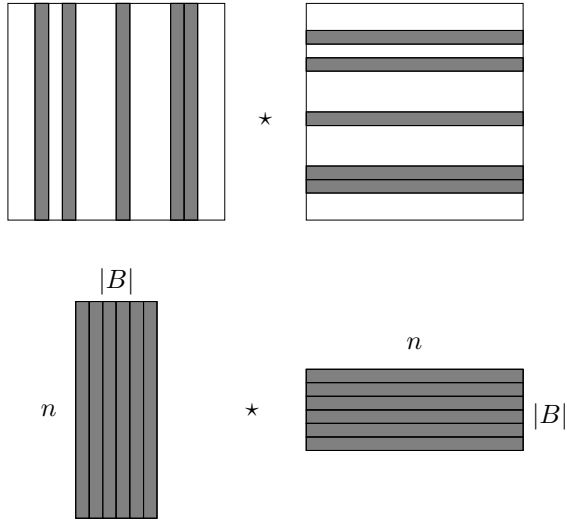

Algorithm \RANDSHORTPATH\ starts by letting $F\gets D$. 
The algorithm then performs $\lceil\log_{3/2}n\rceil$ iterations. In the
$\ell$-th iteration it lets $s\gets (3/2)^\ell$. It then uses a function
called \RAND\ to produce a random subset~$B$ of $V=\{1,2,\ldots,n\}$
obtained by selecting each element of~$V$ independently with probability
$p=(9\ln n)/s$. If $p\ge 1$, then \RAND\ returns the set~$V$.  The
algorithm then constructs the matrices $F[*,B]$ and $F[B,*]$. The matrix
$F[*,B]$ is the matrix whose columns are the columns of~$F$ that
correspond to the vertices of~$B$. Similarly, $F[B,*]$ is the matrix
whose rows are the rows of~$F$ that correspond to the vertices of~$B$
(see Definition~\ref{D-samp} and Figure~\ref{F-sample}).  It then
computes the distance product~$F'$ of the matrices $F[*,B]$ and $F[B,*]$
by calling \DISTPROD, putting a cap of~$sM$ on the absolute values of
all the entries that participate in the product. The call also returns a
matrix~$W'$ of witnesses.  Finally, each element of~$F'$ is compared to
the corresponding element of~$F$.  If the element of~$F'$ is smaller,
then it is copied to~$F$ and the corresponding witness from~$W'$ is
copied to~$W$. (By~$b_{w'_{ij}}$ we denote the $w'_{ij}$-th element of
the set~$B$.)

Let $\delta(i,j)$ denote the (weighted) distance from~$i$ to~$j$ in the
graph, i.e., the smallest weight of a directed path going from~$i$
to~$j$. The weight of a path is the sum of the weights of its edges. The
following lemma is easily established:

\begin{lemma}\label{L-simple} 
    At any stage during the operation of \RANDSHORTPATH,
     for every $i,j\in V$, we have: 
\begin{itemize} 
    \item[(i)] $f_{ij}\ge \delta(i,j)$. 
    \item[(ii)] If $w_{ij}=0$ then $f_{ij}=d_{ij}$. Otherwise, $1\le
    w_{ij}\le n$ and
    $f_{ij} \ge f_{i,w_{ij}} + f_{w_{ij},j}$. 
    \item[(iii)] If $\delta(i,j)=\delta(i,k) + \delta(k,j)$ and if in
    the beginning of some iteration we have $f_{ik}=\delta(i,k)$,
    $f_{kj}=\delta(k,j)$, $|f_{ik}|,|f_{kj}|\le sM$ and $k\in B$, then
    at the end of the iteration we have $f_{ij}=\delta(i,j)$.
\end{itemize}
\end{lemma}

\begin{proof}
    Property $(i)$ clearly holds when~$F$ is
    initialized to~$D$. In each iteration, the algorithm chooses a
    set~$B$ and then lets 
    $$ \begin{array}{l} f'_{ij} \gets \min\{\, f_{ik}+f_{kj} \mid k\in B\,,\,
    |f_{ik}|,|f_{kj}| \le sM\,\} \\ f_{ij} \gets \min\{\, f_{ij} \,,\,
    f'_{ij} \,\}\\ \end{array}$$
    for every $i,j\in V$. For every $k$, we have
    $f_{ik}+ f_{kj} \ge \delta(i,k) + \delta(k,j) \ge \delta(i,j)$, as
    follows from the induction hypothesis and the triangle inequality,
    and thus the new value of $f_{ij}$ is again an upper bound on
    $\delta(i,j)$.  
    
    Property~$(ii)$ also follows easily by induction. Initially,
    $f_{ij}=d_{ij}$ and $w_{ij}=0$, for every $i,j\in V$, so the
    condition is satisfied. Whenever~$f_{ij}$ is assigned a new value,
    we have $1\le w_{ij}\le n$ and $f_{ij} = f_{i,w_{ij}} +
    f_{w_{ij},j}$. Until the next time $f_{ij}$ is assigned a value we
    are thus guaranteed to have $f_{ij} \ge f_{i,w_{ij}} + f_{w_{ij},j}$,
    as the value of $f_{ij}$ does not change, while the values of
    $f_{i,w_{ij}}$ and $f_{w_{ij},j}$ may only decrease.
    
    Finally, if the conditions of property $(iii)$ hold, then at the end
    of the iteration we have $$f_{ij} \;\le\; f'_{ij}\;\le\;
    f_{ik}+f_{kj} \eq \delta(i,k)+\delta(k,j) \eq \delta(i,j)\;.$$ 
    As $f_{ij}\ge \delta(i,j)$,
    by property~$(i)$, we get that $f_{ij}=\delta(i,j)$, as required.
\end{proof}

More interesting is the following lemma:

\begin{lemma}\label{L-rand}
    Let $s=(3/2)^\ell$, for some $1\le \ell\le \lceil\log_{3/2}n\rceil$.
    With very high probability, if there is a shortest path from $i$ to
    $j$ in the graph that uses at most $s$ edges then at
    the end of the $\ell$-th iteration we have $f_{ij}=\delta(i,j)$.
\end{lemma}


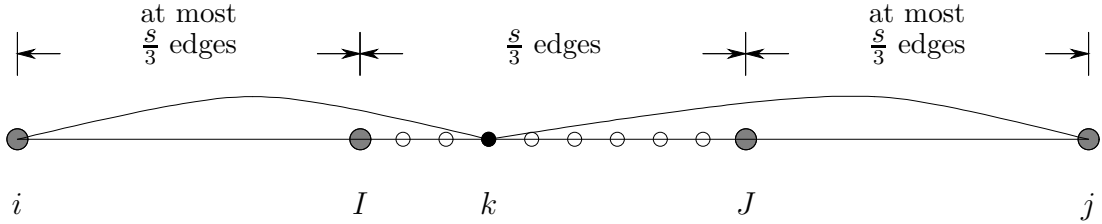
\begin{figure*}[t]
\newcommand{\ls}{\mbox{$s$}}
\newcommand{\AAA}{\shortstack[c]{at most \\[0pt] $\frac{\ls}{3}$ edges}}
\newcommand{\BBB}{$\frac{\ls}{3}$ edges}
\newcommand{\CCC}{\AAA}

$$\setlength{\unitlength}{0.00075000in}
\begingroup\makeatletter\ifx\SetFigFont\undefined%
\gdef\SetFigFont#1#2#3#4#5{%
  \reset@font\fontsize{#1}{#2pt}%
  \fontfamily{#3}\fontseries{#4}\fontshape{#5}%
  \selectfont}%
\fi\endgroup%
{\renewcommand{\dashlinestretch}{30}
\begin{picture}(8207,1341)(0,-10)
\put(3354,564){\ellipse{100}{100}}
\put(3954,564){\ellipse{100}{100}}
\put(4254,564){\ellipse{100}{100}}
\put(4554,564){\ellipse{100}{100}}
\put(4854,564){\ellipse{100}{100}}
\put(5154,564){\ellipse{100}{100}}
\texture{44555555 55aaaaaa aa555555 55aaaaaa aa555555 55aaaaaa aa555555 55aaaaaa 
	aa555555 55aaaaaa aa555555 55aaaaaa aa555555 55aaaaaa aa555555 55aaaaaa 
	aa555555 55aaaaaa aa555555 55aaaaaa aa555555 55aaaaaa aa555555 55aaaaaa 
	aa555555 55aaaaaa aa555555 55aaaaaa aa555555 55aaaaaa aa555555 55aaaaaa }
\put(5454,564){\shade\ellipse{150}{150}}
\put(5454,564){\ellipse{150}{150}}
\put(2754,564){\shade\ellipse{150}{150}}
\put(2754,564){\ellipse{150}{150}}
\put(3054,564){\ellipse{100}{100}}
\put(3654,564){\blacken\ellipse{100}{100}}
\put(3654,564){\ellipse{100}{100}}
\put(7854,564){\shade\ellipse{150}{150}}
\put(7854,564){\ellipse{150}{150}}
\put(354,564){\shade\ellipse{150}{150}}
\put(354,564){\ellipse{150}{150}}
\path(354,564)(7854,564)
\texture{aaffffff ffaaaaaa aaffffff ffaaaaaa aaffffff ffaaaaaa aaffffff ffaaaaaa 
	aaffffff ffaaaaaa aaffffff ffaaaaaa aaffffff ffaaaaaa aaffffff ffaaaaaa 
	aaffffff ffaaaaaa aaffffff ffaaaaaa aaffffff ffaaaaaa aaffffff ffaaaaaa 
	aaffffff ffaaaaaa aaffffff ffaaaaaa aaffffff ffaaaaaa aaffffff ffaaaaaa }
\path(2754,1314)(2754,1014)
\path(2754,1314)(2754,1014)
\path(2754,1164)(3054,1164)
\path(2754,1164)(3054,1164)
\blacken\path(2874.000,1194.000)(2754.000,1164.000)(2874.000,1134.000)(2838.000,1164.000)(2874.000,1194.000)
\path(2454,1164)(2754,1164)
\path(2454,1164)(2754,1164)
\blacken\path(2634.000,1134.000)(2754.000,1164.000)(2634.000,1194.000)(2670.000,1164.000)(2634.000,1134.000)
\path(354,1314)(354,1014)
\path(354,1314)(354,1014)
\path(354,1164)(654,1164)
\path(354,1164)(654,1164)
\blacken\path(474.000,1194.000)(354.000,1164.000)(474.000,1134.000)(438.000,1164.000)(474.000,1194.000)
\path(5454,1314)(5454,1014)
\path(5454,1314)(5454,1014)
\path(5154,1164)(5454,1164)
\path(5154,1164)(5454,1164)
\blacken\path(5334.000,1134.000)(5454.000,1164.000)(5334.000,1194.000)(5370.000,1164.000)(5334.000,1134.000)
\path(5454,1164)(5754,1164)
\path(5454,1164)(5754,1164)
\blacken\path(5574.000,1194.000)(5454.000,1164.000)(5574.000,1134.000)(5538.000,1164.000)(5574.000,1194.000)
\path(7854,1314)(7854,1014)
\path(7854,1314)(7854,1014)
\path(7554,1164)(7854,1164)
\path(7554,1164)(7854,1164)
\blacken\path(7734.000,1134.000)(7854.000,1164.000)(7734.000,1194.000)(7770.000,1164.000)(7734.000,1134.000)
\path(3654,564)	(3725.927,575.392)
	(3796.591,586.536)
	(3866.006,597.433)
	(3934.190,608.085)
	(4001.156,618.494)
	(4066.920,628.661)
	(4131.499,638.589)
	(4194.906,648.279)
	(4257.159,657.733)
	(4318.271,666.952)
	(4378.260,675.938)
	(4437.139,684.693)
	(4494.925,693.218)
	(4551.633,701.516)
	(4607.278,709.588)
	(4661.876,717.435)
	(4715.443,725.060)
	(4767.993,732.464)
	(4819.543,739.649)
	(4870.107,746.616)
	(4919.701,753.368)
	(4968.341,759.906)
	(5016.042,766.231)
	(5062.820,772.346)
	(5108.689,778.253)
	(5153.666,783.952)
	(5197.766,789.445)
	(5241.004,794.735)
	(5283.396,799.823)
	(5324.956,804.711)
	(5405.647,813.893)
	(5483.201,822.294)
	(5557.740,829.928)
	(5629.388,836.810)
	(5698.269,842.952)
	(5764.507,848.369)
	(5828.225,853.074)
	(5889.546,857.081)
	(5948.595,860.404)
	(6005.495,863.056)
	(6060.369,865.051)
	(6113.341,866.404)
	(6164.535,867.127)
	(6214.075,867.234)
	(6262.083,866.740)
	(6308.683,865.657)
	(6354.000,864.000)

\path(6354,864)	(6405.697,861.055)
	(6460.438,856.515)
	(6518.772,850.271)
	(6581.248,842.212)
	(6648.416,832.230)
	(6720.824,820.213)
	(6799.022,806.052)
	(6840.464,798.133)
	(6883.560,789.637)
	(6928.378,780.551)
	(6974.987,770.859)
	(7023.455,760.549)
	(7073.851,749.607)
	(7126.244,738.020)
	(7180.703,725.772)
	(7237.296,712.852)
	(7296.092,699.244)
	(7357.159,684.935)
	(7420.567,669.912)
	(7486.383,654.161)
	(7554.677,637.668)
	(7625.517,620.419)
	(7698.971,602.400)
	(7775.110,583.599)
	(7854.000,564.000)

\path(354,564)	(397.668,574.437)
	(440.574,584.657)
	(482.726,594.660)
	(524.134,604.448)
	(604.755,623.388)
	(682.513,641.490)
	(757.484,658.769)
	(829.743,675.237)
	(899.365,690.908)
	(966.427,705.797)
	(1031.003,719.917)
	(1093.169,733.281)
	(1153.000,745.904)
	(1210.573,757.799)
	(1265.963,768.980)
	(1319.245,779.460)
	(1370.494,789.254)
	(1419.787,798.375)
	(1467.199,806.837)
	(1512.806,814.653)
	(1556.682,821.838)
	(1598.903,828.404)
	(1678.685,839.739)
	(1752.755,848.766)
	(1821.717,855.595)
	(1886.176,860.338)
	(1946.735,863.103)
	(2004.000,864.000)

\path(2004,864)	(2061.265,863.103)
	(2121.824,860.338)
	(2186.283,855.595)
	(2255.245,848.766)
	(2329.315,839.739)
	(2409.097,828.404)
	(2451.318,821.838)
	(2495.194,814.653)
	(2540.801,806.837)
	(2588.212,798.375)
	(2637.506,789.254)
	(2688.755,779.460)
	(2742.037,768.980)
	(2797.427,757.799)
	(2855.000,745.904)
	(2914.831,733.281)
	(2976.997,719.917)
	(3041.573,705.797)
	(3108.635,690.908)
	(3178.257,675.237)
	(3250.516,658.769)
	(3325.487,641.490)
	(3403.245,623.388)
	(3483.866,604.448)
	(3525.274,594.660)
	(3567.426,584.657)
	(3610.332,574.437)
	(3654.000,564.000)

\put(1554,1164){\makebox(0,0)[b]{\smash{{{\SetFigFont{11}{13.2}{\rmdefault}{\mddefault}{\updefault}\AAA}}}}}
\put(2754,39){\makebox(0,0)[b]{\smash{{{\SetFigFont{11}{13.2}{\rmdefault}{\mddefault}{\updefault}\large $I$}}}}}
\put(5454,39){\makebox(0,0)[b]{\smash{{{\SetFigFont{11}{13.2}{\rmdefault}{\mddefault}{\updefault}\large $J$}}}}}
\put(3654,39){\makebox(0,0)[b]{\smash{{{\SetFigFont{11}{13.2}{\rmdefault}{\mddefault}{\updefault}\large $k$}}}}}
\put(354,39){\makebox(0,0)[b]{\smash{{{\SetFigFont{11}{13.2}{\rmdefault}{\mddefault}{\updefault}\large $i$}}}}}
\put(7854,39){\makebox(0,0)[b]{\smash{{{\SetFigFont{11}{13.2}{\rmdefault}{\mddefault}{\updefault}\large $j$}}}}}
\put(4104,1164){\makebox(0,0)[b]{\smash{{{\SetFigFont{11}{13.2}{\rmdefault}{\mddefault}{\updefault}\BBB}}}}}
\put(6654,1164){\makebox(0,0)[b]{\smash{{{\SetFigFont{11}{13.2}{\rmdefault}{\mddefault}{\updefault}\CCC}}}}}
\end{picture}
}$$ 
\caption{\label{F-cor}The correctness proof of \RANDSHORTPATH.}
\end{figure*}

\begin{proof}
    We prove the lemma by induction of~$\ell$. It is easy to check that
    the claim holds for $\ell=1$. We show next that if the
    claim holds for $\ell-1$, then it also holds for~$\ell$.  Let~$i$
    and~$j$ be two vertices connected by a shortest path that uses at
    most $s=(3/2)^\ell$ edges. Let~$p$ be such a shortest path from $i$
    to~$j$. If the number of edges on~$p$ is at most~$2s/3$ then, by the
    induction hypothesis, after the $(\ell-1)$-st iteration we already
    have $f_{ij}=\delta(i,j)$ (with very high probability). Suppose,
    therefore, that the number of edges on~$p$ is at least~$2s/3$ and at
    most~$s$. To avoid technicalities, we `pretend' at first that $s/3$
    is an integer. We later indicate the changes needed to make the
    proof rigorous.
    
    Let~$I$ and~$J$ be vertices on~$p$ such that~$I$ and~$J$ are
    separated, on~$p$, by {\em exactly\/} $s/3$ edges, and such that~$i$
    and~$I$, and~$J$ and~$j$ are separated, on~$p$, by {\em at most\/} $s/3$
    edges. See Figure~\ref{F-cor}. Such vertices $I$ and $J$ can always
    be found as the path~$p$ is composed of at least $2s/3$ and at
    most~$s$ edges.
     
    Let~$A$ be the set of vertices lying between~$I$ and~$J$
    (inclusive) on~$p$. Note that $|A|\ge s/3$. Let $k\in A$. As~$k$
    lies on a shortest path from~$i$ to~$j$, we have
    $\delta(i,j)=\delta(i,k)+\delta(k,j)$. As~$k$ lies between~$I$ and
    $J$, there are shortest paths from~$i$ to~$k$, and from~$k$ to~$j$
    that use at most $2s/3$ edges. By the induction hypothesis, we get
    that at the beginning of the $\ell$-th iteration we have
    $f_{ik}=\delta(i,k)$ and $f_{kj}=\delta(k,j)$, with very high
    probability. We also have $|f_{ik}|,|f_{kj}|\le sM$. It follows,
    therefore, from Lemma~\ref{L-simple}$(iii)$, that if there exists
    $k\in A\cap B$, where~$B$ is the set of vertices chosen at the
    $\ell$-th iteration, then at the end of the $\ell$-th iteration we
    have $f_{ij}=\delta(i,j)$, as required.
    
    What is the probability that $A\cap B\ne \phi$?
    Let $p=(9\ln n)/s$. If $p\ge 1$, then $B=V$ and clearly $A\cap
    B\ne\phi$. Suppose, therefore, that $p=(9\ln n)/s<1$. Each vertex
    then belongs to $B$ independently with probability~$p$.
    As $|A|\ge s/3$, the probability that $A\cap B=\phi$ is at most
    $$\left(1-\frac{9\ln n}{s}\right)^{s/3}\le {\rm e}^{-3\ln n} =
    n^{-3}\;.$$
    
    As there are less than $n^2$ pairs of vertices in the graph, the
    probability of failure during the entire operation of the algorithm
    is at most $n^2\cdot n^{-3} = 1/n$. (We do not have to multiply the
    probability by the number of iterations, as 
    each pair of vertices should only be considered at one of the
    iterations. If a pair $i,j\in V$ violates the
    condition of the lemma, then it also does so at the $\ell$-th
    iteration, where $\ell$ is the smallest integer such that there is a
    shortest path from $i$ to $j$ that uses at most $s=(3/2)^\ell$
    edges.)
    
    Unfortunately, $s/3$ is not an integer. To make the proof go
    through, we prove by induction a slight strengthening of the lemma.
    Define the sequence $s_0=1$ and $s_\ell = \lceil 3s_{\ell-1}/2
    \rceil$, for $\ell>0$. Note that $s_\ell\ge (3/2)^\ell$. We show by
    induction on~$\ell$ that, with high probability, for every $i,j\in
    V$, if there is a shortest path from~$i$ to~$j$ that uses at
    most~$s_\ell$ edges, then at the end of the $\ell$-th iteration we
    have $f_{ij}=\delta(i,j)$. The proof is almost the same as before.
    If~$p$ is a shortest path from~$i$ to~$j$ that uses at most $s_\ell$
    edges, we consider vertices~$I$ and~$J$ on~$p$ such that~$I$ and~$J$
    are separated by exactly $\lfloor s_\ell/2 \rfloor$ edges, and such
    that~$i$ and~$I$, and~$J$ and~$j$ are separated by at most $\lceil
    s_\ell/2 \rceil$ edges.  Repeating the above arguments we obtain a
    rigorous proof of the (strengthened) lemma.
\end{proof}

Combining Lemmas~\ref{L-simple} and~\ref{L-rand} with the fact that each
pair of vertices in a graph of~$n$ vertices is connected by a shortest
path that uses less than~$n$ edges, assuming there are no negative
cycles in the graph, we get that after the last iteration, $F$ is, with
very high probability, the distance matrix of the graph. Furthermore,
either $\delta(i,j)=d_{ij}$, or $w_{ij}$ lies on a shortest path
from~$i$ to~$j$. This is stated formally in the following lemma:

\begin{lemma}\label{L-correct}
    If there are no negative weight cycles in the graph, then after the
    last iteration of \RANDSHORTPATH, with very high probability, for
    every $i,j\in V$ we have \vspace*{-5pt}
\begin{itemize}
\item[(i)] $f_{ij}=\delta(i,j)$.
\item[(ii)] If $w_{ij}=0$ then $\delta(i,j)=d_{ij}$. Otherwise,
$1\le w_{ij}\le n$ and $\delta(i,j)=\delta(i,w_{ij})+\delta(w_{ij},j)$.
\end{itemize}
\end{lemma}

\begin{proof} Condition~$(i)$ follows, as mentioned, from
    Lemma~\ref{L-rand}, the fact that in the last iteration $s\ge n$,
    and the fact that if $\delta(i,j)<+\infty$, and if there are no
    negative weight cycles in the graph, then there is a shortest path
    from~$i$ to~$j$ that uses at most~$n-1$ edges. 

    Suppose now that $f_{ij}=\del(i,j)<d_{ij}$. By Lemma~\ref{L-simple}$(ii)$
    we get that after the last iteration we have $1\le w_{ij}\le n$ and 
    $f_{ij} \ge f_{i,w_{ij}} + f_{w_{ij},j}$, or equivalently,
    $\del(i,j)\ge \del(i,w_{ij}) + \del(w_{ij},j)$. But, by the triangle
    inequality we have $\del(i,j)\le \del(i,w_{ij}) +
    \del(w_{ij},j)$. Thus, $\del(i,j)= \del(i,w_{ij}) +
    \del(w_{ij},j)$, as required.
\end{proof}

It is also easy to see that the input graph contains a negative cycle if
and only if $f_{ii}<0$ for some $1\le i\le n$. If there is a path from~$i$
to~$j$ that passes though a vertex contained in a negative cycle, we define
the distance from $i$ to $j$ to be $-\infty$. Using a standard method, it
is easy to identify all such pairs in $\Ot(n^\omega)$
time. See Galil and Margalit \cite{GaMa97b} for the details. 

The matrix~$W$ returned by \RANDSHORTPATH\ contains a succinct
representation of shortest paths between all pairs of vertices in the
graph. Ways for reconstructing these shortest paths are
described in the next section.

\begin{figure}
\begin{center}
\epsfxsize=2.3in
\makebox[0pt]{\epsfbox{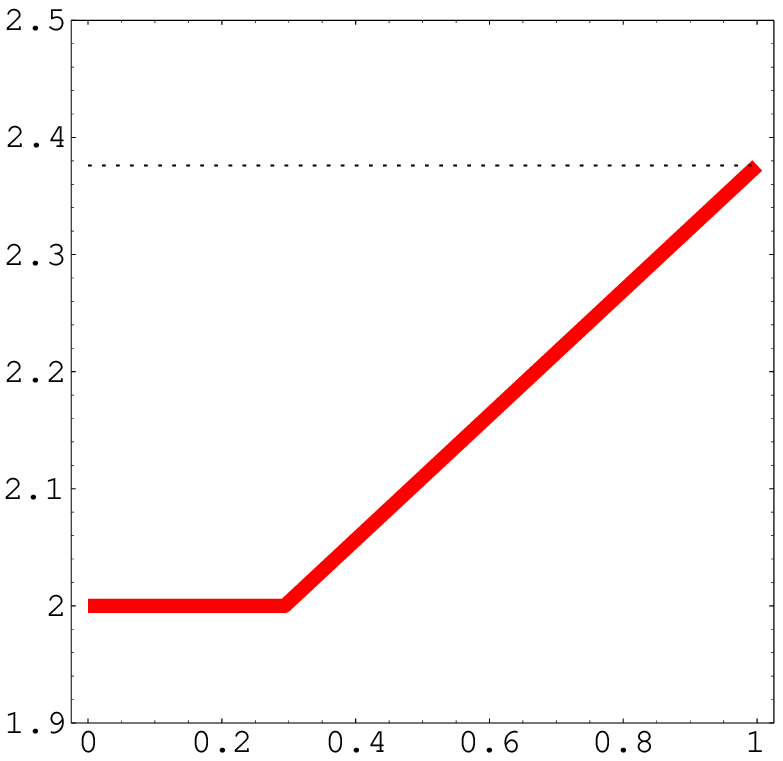} $\quad$ 
\epsfxsize=2.3in
\epsfbox{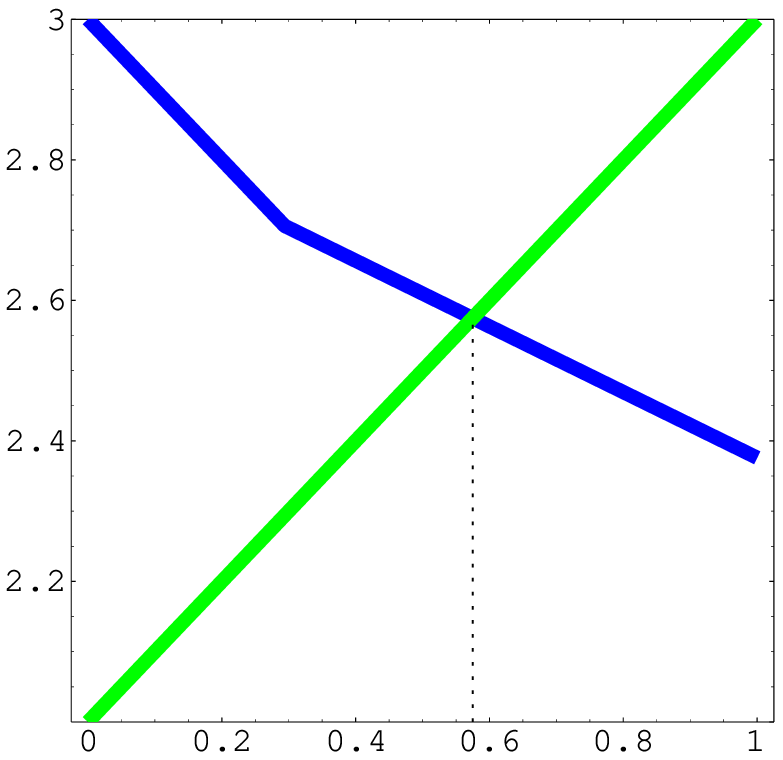} } \\[15pt]
\makebox[0pt]{
\makebox[2.3in][c]{$\omega(1,r,1)$}
$\quad$
\makebox[2.3in][c]{
$\omega(1,r,1)+(1-r)$ and $2+r$}}
\end{center}
\caption{Best available bounds on the functions $\omega(1,r,1)$ and 
$\omega(1,r,1)+(1-r)$, and the function $2+r$.}
\label{F-omega}
\end{figure}

What is the complexity of \RANDSHORTPATH? The time taken by the
$\ell$-th iteration is dominated by the time needed to compute the distance
product of an $n\times m$ matrix by an $m\times n$ matrix, where
$m=O((n\log n)/s)$, with entries of absolute value at most~$sM$ using
\DISTPROD. If we assume that $s=n^{1-r}$ and $M=n^t$, then according to 
Lemma~\ref{L-dist-prod}, this time is
$\Ot(\min\{n^{t+\omega(1,r,1)+(1-r)},n^{2+r}\})$. Graphs of the best
available upper bounds on the functions $\omega(1,r,1)$ and
$\omega(1,r,1)+(1-r)$ are given in Figure~\ref{F-omega}. (Also shown
there is the function $2+r$.)  Note that
$\omega(1,r,1)+(1-r)$ is decreasing in~$r$ while $2+r$ is
increasing in~$r$.
The running time of an iteration is
maximized when $t+\omega(1,r,1)+(1-r) = 2+r$, or equivalently, when
$\omega(1,r,1)=1+2r-t$. As there are only $O(\log n)$ iterations, we get

\begin{theorem}
    Algorithm \RANDSHORTPATH\ finds, with very high probability, all
    distances in the input graph, and a succinct representation of
    shortest paths between all pairs of vertices in the graph.  If the
    input graph has~$n$ vertices, and the weights are all integers with
    absolute values at most $M=n^t$, where $t\le 3-\omega$, then its
    running time is $\Ot(n^{2+\mu(t)})$, where $\mu=\mu(t)$ satisfies
    $\omega(1,\mu,1)=1+2\mu-t$.
\end{theorem}

If $M>n^{3-\omega}$ then fast matrix multiplication algorithms are
never used by the algorithm and the running time is then $\Ot(n^3)$.

Let us look more closely at the running time of the algorithm when $M=O(1)$.
This is the case, for example, if all the weights in the graph belong to
the set $\{-1,0,1\}$. The running time of the algorithm of Alon, Galil and
Margalit in this case is $\Ot(n^{(3+\omega)/2})$, which is about
$O(n^{2.688})$. The running time of the new algorithm is $\Ot(n^{2+\mu})$,
where~$\mu$ satisfies $\omega(1,\mu,1)=1+2\mu$. Using the naive bound
$\omega(1,r,1)\le 2+(\omega-2)r$, we get that $\mu\le
\frac{1}{4-\omega}<0.616$. Using the improved bound of Lemma~\ref{L-omega},
we get that $\mu\le \frac{\alpha(\omega-1)-1}{\omega+2\alpha-4}<
0.575$.

\begin{corol}
    Algorithm \RANDSHORTPATH\ finds, with very high probability, all
    distances, and a succinct representation of shortest paths between
    all pairs of vertices in the graph on~$n$ vertices in
    which all the weights are taken from the set $\{-1,0,1\}$ in
    $O(n^{2.575})$ time.
\end{corol}

\section{Constructing shortest paths}
\label{S-paths}

A simple recursive algorithm, \PATH, for constructing shortest paths is
given in Figure~\ref{F-path}. If there are no negative weight cycles in
the graph, and if~$W$ is the matrix of witnesses returned by a
successful run of \RANDSHORTPATH, then $\PATH(W,i,j)$ returns a shortest
path from~$i$ to~$j$ in the graph. If $w_{ij}=0$, then the edge $(i,j)$
is a shortest path from~$i$ to~$j$. Otherwise, a shortest path from~$i$
to~$j$ is obtained by concatenating a shortest path from~$i$
to~$w_{ij}$, found using a recursive call to \PATH, and a shortest path
from~$w_{ij}$ to~$j$, found using a second recursive call to \PATH. (The
dot in next to last line in the description of \PATH\ is used to denote
concatenation.) If there is no directed path from~$i$ to~$j$ in the
graph, then $\PATH(W,i,j)$ returns the ``edge'' $(i,j)$ whose weight is
$+\infty$.

\begin{figure}[t]
\begin{center}
\null\quad\\
\framebox{\hspace{0.4cm}\parbox{4.8in}{
\algorithm\ \PATH$(W,i,j)$ $\phantom{2^{2^{2^2}}}$ \\[5pt]
{\tt if} $w_{ij}=0$ {\tt then} \\
$\null\qquad${\tt return} $\langle i,j\rangle$\\
{\tt else} \\
$\null\qquad${\tt return}
$\PATH(W,i,w_{ij})\,.\,\PATH(W,w_{ij},j)$ \\
{\tt endif}$\phantom{2_{2_{2_2}}}$
}\hspace{0.6cm}}
 \end{center}
\caption{\label{F-path}Constructing a shortest path using a matrix of
   witnesses.}
\end{figure}

\begin{theorem}\label{T-path}
    If there are no negative weight cycles in the input graph, and
    if~$W$ is the matrix of witnesses returned by a successful run of
    \RANDSHORTPATH, then $\PATH(W,i,j)$ returns a shortest path from~$i$
    to~$j$ in the graph. The running time of $\PATH(W,i,j)$ is
    proportional to the number of edges in the path returned.
\end{theorem}

\begin{proof}
    For every $i,j\in V$, let~$t_{ij}$ be the number of the iteration of
    \RANDSHORTPATH\ in which~$f_{ij}$ was set for the last time. If
    $f_{ij}=d_{ij}$, let $t_{ij}=0$. We need the following claim:

\begin{claim}\label{C-times}
If $1\le w_{ij}\le n$, then $t_{i,w_{ij}},t_{w_{ij},j}<t_{ij}$.
\end{claim}

\begin{proof}
Suppose that $f_{ij}$ was set for the last time at the $\ell$-th
iteration. Let $f^0_{rs}$ be the elements of the matrix $F$ at the
beginning of the $\ell$-th iteration, and $f^1_{rs}$ be these elements
at the end of the $\ell$-th iteration. By our assumption and by
Lemma~\ref{L-correct}, we get that
$$\begin{array}{c}
f_{ij}\eq f^1_{ij} \eq f^0_{i,w_{ij}} + f^0_{w_{ij},j}\;,\\
f_{ij}\eq \del(i,j) = \del(i,w_{ij}) + \del(w_{ij},j)\;.\\
\end{array}$$
As $f^0_{i,w_{ij}}\ge \del(i,w_{ij})$ and $f^0_{w_{ij},j} \ge
\del(w_{ij},j)$ (see Lemma~\ref{L-simple}$(i)$), we get that
$f^0_{i,w_{ij}}=\del(i,w_{ij})$ and $f^0_{w_{ij},j} =
\del(w_{ij},j)$. Thus, $f_{i,w_{ij}}$ and $f_{w_{ij},j}$ are already
assigned their final values at the beginning of the $\ell$-th iteration,
and therefore $t_{i,w_{ij}},t_{w_{ij},j}<\ell=t_{ij}$, as required.
\end{proof}

We now prove Theorem~\ref{T-path} by induction on~$t_{ij}$. If
$t_{ij}=0$, then $w_{ij}=0$, and $\PATH(W,i,j)$ returns the edge $(i,j)$ which
is indeed a shortest path from~$i$ to~$j$. Suppose now that
$\PATH(W,i,j)$ returns a shortest path from~$r$ to~$s$ for every~$r$
and~$s$ for which $t_{rs}<\ell$. Suppose that $t_{ij}=\ell$. By
Claim~\ref{C-times}, we get that $t_{i,w_{ij}}, t_{w_{ij},j}<\ell$. By
the induction hypothesis, the recursive calls $\PATH(W,i,w_{ij})$ and
$\PATH(W,w_{ij},j)$ return shortest paths from~$i$ to~$w_{ij}$ and
from~$w_{ij}$ to~$j$. As $\del(i,j) = \del(i,w_{ij}) + \del(w_{ij},j)$
(Lemma~\ref{L-correct}), the concatenation of these two shortest paths
is indeed a shortest path from~$i$ to~$j$, as required.
\end{proof}

There is, however, something unsatisfying with the behavior of $\PATH$.
While it is true that the call $\PATH(W,i,j)$ always returns a shortest
path from~$i$ to~$j$ in the graph, the shortest path returned is not
necessarily {\em simple}, i.e., it may visit certain vertices more than
once. This may happen, of course, only if there are zero weight cycles
in the graph. It is, of course, easy to convert a non-simple shortest
path into a simple shortest path, by removing cycles, but the running
time then is no longer proportional to the number of edges on the
shortest path produced.

Another possible objection to the use of \PATH\ is that it cannot
produce shortest paths in {\em real time}. While it is
true that a shortest path that uses $\ell$ edges can be found in
$O(\ell)$ time, it may also take $\Omega(\ell)$ time just to find the
second vertex on such a path.

\begin{figure}[t]
\begin{center}
\null\quad\\
\framebox{\hspace{0.4cm}\parbox{4.8in}{
\algorithm\ \SPATH$(S,i,j)$ $\phantom{2^{2^{2^2}}}$ \\[5pt]
{\tt if} $s_{ij}=j$ {\tt then} \\
$\null\qquad${\tt return} $\langle i,j\rangle$\\
{\tt else} \\
$\null\qquad${\tt return}
$\langle i,s_{ij}\rangle\,.\,\SPATH(S,s_{ij},j)$ \\
{\tt endif}$\phantom{2_{2_{2_2}}}$
}\hspace{0.6cm}}
 \end{center}
\caption{\label{F-SPATH}Constructing a shortest path using a matrix of
 successors.}
\end{figure}

\begin{figure}[t]
\begin{center}
\null\quad\\
\framebox{\hspace{0.4cm}\parbox{4.8in}{
\algorithm\ \WITTOSUC$(W,T)$ $\phantom{2^{2^{2^2}}}$ \\[5pt]
$S\gets 0$ \\[5pt]
{\tt for} $\ell\gets 0$ {\tt to} $\max(T)$ {\tt do}
$T_\ell\gets\{\,(i,j)\mid t_{ij}=\ell\,\}$ \\[5pt]
{\tt for every} $(i,j)\in T_0$ {\tt do} $s_{ij}\gets j$\\[5pt]
{\tt for} $\ell\gets 1$ {\tt to} $\max(T)$ {\tt do} \\
\null{\tt for every} $(i,j)\in T_\ell$ {\tt do} \\
\null{\tt begin}\\
\null\qquad$k\gets w_{ij}$\\
\null\qquad{\tt while} $s_{ij}=0$ {\tt do} 
$s_{ij}\gets s_{ik}$ ; $j\gets s_{ij}$ \\
\null{\tt end}\\[5pt]
{\tt return} $S$ $\phantom{2_{2_{2_2}}}$
}\hspace{0.6cm}}
 \end{center}
\caption{\label{F-succ}Constructing a matrix of successors.}
\end{figure}

To address these two issues, we show next that the matrix of
witnesses~$W$ returned by \RANDSHORTPATH\ can be easily converted into a
matrix of {\em successors\/} (see, e.g., \cite{CoLeRi90}, Chapter~25,
were predecessors, instead of successors are considered). A matrix of
successors can be easily used to construct trees of shortest paths.

\begin{definition}[Successors]
    A matrix~$S$ is a matrix of successors for a graph $G=(V,E)$ if for
    every $i,j\in V$, if there is a path from~$i$ to~$j$ in the graph,
    then the call $\SPATH(S,i,j)$, where \SPATH\ is the procedure given
    in Figure~\ref{F-SPATH}, returns a {\em simple\/} shortest path
    from~$i$ to~$j$ in the graph.
\end{definition}

Algorithm \WITTOSUC, given in Figure~\ref{F-succ}, receives a matrix~$W$
of witnesses returned by \RANDSHORTPATH, and a matrix~$T$ that gives the
iteration number in which each element of~$W$ was set for the last time,
as in the proof of Theorem~\ref{T-path}. (It is very easy, of course,
to modify \RANDSHORTPATH\ so that it would also return this matrix.) It
returns a matrix~$S$ of successors. Algorithm \WITTOSUC\ works correctly
even if there are zero weight cycles in the graph, but not if there are
negative weight cycles in the graphs as then distances and shortest
paths are not well defined.

\begin{theorem}
    If there are no negative weight cycles in the graph, if~$W$ is 
    the matrix of witnesses returned by a successful run of
    \RANDSHORTPATH, and if~$T$ is the corresponding matrix of iteration
    numbers, then algorithm \WITTOSUC\ returns a matrix of successors.
    The running time of algorithm \WITTOSUC\ is $O(n^2)$.
\end{theorem}

\begin{proof} 
    Algorithm \WITTOSUC\ begins by initializing all the elements of the
    $n\times n$ matrix~$S$ to~0. It then constructs, for each iteration
    number~$\ell$, the set~$T_\ell$ of pairs $(i,j)$ for which
    $t_{ij}=\ell$. It is easy to construct all these sets in~$O(n^2)$
    by bucket sorting. (In the description of \WITTOSUC, $\max(T)$
    denotes the maximal element in~$T$. Note that $\max(T)=O(\log n)$.)
    Next, for every $(i,j)$ such that $t_{ij}=0$, it sets $s_{ij}\gets
    j$. It then performs~$\max(T)$ iterations, one of each iteration of 
    \RANDSHORTPATH\ in which values are changed. 
    
    We prove, by induction on the order in which the elements of the
    matrix~$S$ are assigned nonzero values, that if $s_{ij}\ne 0$, then
    $\SPATH(S,i,j)$ returns a simple shortest path from~$i$ to~$j$ in
    the graph. This clearly holds after \WITTOSUC\ sets $s_{ij}\gets j$
    for every $(i,j)\in T_0$, as the edge $(i,j)$ is then a simple
    shortest path from~$i$ to~$j$ in the graph.
    
    Suppose that \WITTOSUC\ is now about to perform the {\tt while} loop
    for a pair $(i,j)$ for which $t_{ij}=\ell$. If $s_{ij}\ne 0$, then
    no new entries are assigned nonzero values. Suppose, therefore, that
    $s_{ij}=0$. Let $k=w_{ij}$. By Claim~\ref{C-times}, we get that
    $t_{ik}<\ell$ and $,t_{kj}<\ell$. Thus, $s_{ik}$ and $s_{kj}$ are
    already assigned nonzero values and by the induction hypothesis, the
    calls $\SPATH(S,i,k)$ and $\SPATH(S,k,j)$ return simple
    shortest paths in the graph from~$i$ to~$k$, and from~$k$ to~$j$.
    Let~$v$ be the first vertex on the path $\SPATH(S,i,k)$ for which
    $s_{vj}\ne 0$. The vertex~$v$ is well defined as $s_{kj}\ne 0$. As
    $s_{vj}\ne 0$, we get, by the induction hypothesis, that
    $\SPATH(S,v,j)$ traces a simple shortest path from~$v$ to~$j$. The
    concatenation of the portion of $\SPATH(S,i,k)$ from~$i$ to~$v$, and
    of $\SPATH(S,v,j)$ is clearly a shortest path from~$i$
    to~$j$. It is also simple as both portions are simple, and as for
    every~$u$ on the first portion, except~$v$, we have $s_{uj}=0$,
    while for every~$u$ on the second portion we have $s_{uj}\ne
    0$. After the {\tt while} loop corresponding to~$(i,j)$,
    $\SPATH(S,i,j)$ returns this simple shortest path. Furthermore,
    if $s_{uj}$ is changed by this {\tt while} loop, then~$u$ lies on the
    first portion of this simple shortest path, and $\SPATH(S,u,j)$
    is the corresponding suffix of this simple shortest path,
    which is also a simple shortest path.
    
    Finally, the complexity of the algorithm is $O(n^2)$ as each
    iteration of the {\tt while} loop reduces the number of zero
    elements of~$S$ by one.
\end{proof}

\section{A deterministic algorithm for unweighted graphs} 
\label{S-deter}

In this section we describe a deterministic version of algorithm
\RANDSHORTPATH\ of Section~\ref{S-exact}. The version described here
works only for {\em unweighted\/} directed graphs. A Slightly more
complicated deterministic algorithm that works for weighted directed
graphs is described in the next section. We start with the following
useful definition:

\begin{definition}[$\del(i,j)$ and $\eta(i,j)$]
As before, let $\delta(i,j)$ denote the distance
from~$i$ to~$j$ in the graph, i.e., the minimum weight of a path
from~$i$ to~$j$ in the graph, where the weight of a path is the sum of
the weights of its edges. Let $\eta(i,j)$ denote the minimum
{\em number of edges\/} on a shortest path from~$i$ to~$j$. 
\end{definition}

If the graph is unweighted then $\delta(i,j)=\eta(i,j)$, for every
$i,j\in V$. Note that $\eta(i,j)$ is not necessarily the distance
from~$i$ to~$j$ in the unweighted version of the graph.

Algorithm \RANDSHORTPATH\ implicitly used the notion of {\em bridging
   sets\/} which we now formalize:

\begin{definition}[Bridging sets]
    Let $G=(V,E)$ be a weighted directed graph and let $s\ge 1$.
    A set of vertices~$B$ is said to be an {\em $s$-bridging
       set\/} if for every two vertices $i,j\in V$ such that
    $\eta(i,j)\ge s$, i.e., if all shortest paths from $i$ to $j$ use at
    least $s$ edges, there exists $k\in B$, such that
    $\delta(i,j)=\delta(i,k)+\delta(k,j)$. The set~$B$ is said to be a
    {\em strong $s$-bridging set\/} if for every two vertices $i,j\in V$
    such that $\eta(i,j)\ge s$, there exists $k\in B$, such
    that $\delta(i,j)=\delta(i,k)+\delta(k,j)$ and
    $\eta(i,j)=\eta(i,k)+\eta(k,j)$.
\end{definition}

\begin{figure}[t]
$$\setlength{\unitlength}{0.00055in} 
\begingroup\makeatletter\ifx\SetFigFont\undefined
\def\x#1#2#3#4#5#6#7\relax{\def\x{#1#2#3#4#5#6}}%
\expandafter\x\fmtname xxxxxx\relax \def\y{splain}%
\ifx\x\y   
\gdef\SetFigFont#1#2#3{%
  \ifnum #1<17\tiny\else \ifnum #1<20\small\else
  \ifnum #1<24\normalsize\else \ifnum #1<29\large\else
  \ifnum #1<34\Large\else \ifnum #1<41\LARGE\else
     \huge\fi\fi\fi\fi\fi\fi
  \csname #3\endcsname}%
\else
\gdef\SetFigFont#1#2#3{\begingroup
  \count@#1\relax \ifnum 25<\count@\count@25\fi
  \def\x{\endgroup\@setsize\SetFigFont{#2pt}}%
  \expandafter\x
    \csname \romannumeral\the\count@ pt\expandafter\endcsname
    \csname @\romannumeral\the\count@ pt\endcsname
  \csname #3\endcsname}%
\fi
\fi\endgroup
{\renewcommand{\dashlinestretch}{30}
\begin{picture}(4607,2906)(0,-10)
\texture{44555555 55aaaaaa aa555555 55aaaaaa aa555555 55aaaaaa aa555555 55aaaaaa 
	aa555555 55aaaaaa aa555555 55aaaaaa aa555555 55aaaaaa aa555555 55aaaaaa 
	aa555555 55aaaaaa aa555555 55aaaaaa aa555555 55aaaaaa aa555555 55aaaaaa 
	aa555555 55aaaaaa aa555555 55aaaaaa aa555555 55aaaaaa aa555555 55aaaaaa }%
\put(354,2510){\shade\ellipse{212}{212}}
\put(354,2510){\ellipse{212}{212}}
\put(4254,2510){\shade\ellipse{212}{212}}
\put(4254,2510){\ellipse{212}{212}}
\put(1554,2285){\shade\ellipse{150}{150}}
\put(1554,2285){\ellipse{150}{150}}
\put(354,1310){\shade\ellipse{212}{212}}
\put(354,1310){\ellipse{212}{212}}
\put(4254,1310){\shade\ellipse{212}{212}}
\put(4254,1310){\ellipse{212}{212}}
\put(1854,935){\shade\ellipse{150}{150}}
\put(1854,935){\ellipse{150}{150}}
\path(354,2510)	(405.851,2520.437)
	(456.793,2530.656)
	(555.993,2550.446)
	(651.690,2569.385)
	(743.972,2587.487)
	(832.928,2604.765)
	(918.649,2621.232)
	(1001.223,2636.903)
	(1080.740,2651.791)
	(1157.288,2665.911)
	(1230.958,2679.275)
	(1301.838,2691.897)
	(1370.017,2703.792)
	(1435.586,2714.972)
	(1498.633,2725.453)
	(1559.247,2735.247)
	(1617.518,2744.367)
	(1673.534,2752.829)
	(1727.386,2760.646)
	(1779.163,2767.830)
	(1828.953,2774.397)
	(1922.932,2785.732)
	(2010.037,2794.760)
	(2090.982,2801.591)
	(2166.481,2806.335)
	(2237.249,2809.101)
	(2304.000,2810.000)

\path(2304,2810)	(2370.751,2809.105)
	(2441.519,2806.341)
	(2517.018,2801.600)
	(2597.963,2794.771)
	(2685.068,2785.745)
	(2779.047,2774.411)
	(2828.837,2767.845)
	(2880.614,2760.660)
	(2934.466,2752.844)
	(2990.483,2744.383)
	(3048.753,2735.262)
	(3109.367,2725.468)
	(3172.414,2714.987)
	(3237.983,2703.806)
	(3306.162,2691.911)
	(3377.042,2679.287)
	(3450.712,2665.923)
	(3527.260,2651.803)
	(3606.777,2636.913)
	(3689.351,2621.241)
	(3775.072,2604.772)
	(3864.028,2587.494)
	(3956.310,2569.391)
	(4052.007,2550.450)
	(4151.207,2530.657)
	(4202.149,2520.438)
	(4254.000,2510.000)

\path(354,2510)	(405.851,2499.563)
	(456.793,2489.344)
	(555.993,2469.554)
	(651.690,2450.615)
	(743.972,2432.513)
	(832.928,2415.235)
	(918.649,2398.768)
	(1001.223,2383.097)
	(1080.740,2368.209)
	(1157.288,2354.089)
	(1230.958,2340.725)
	(1301.838,2328.103)
	(1370.017,2316.208)
	(1435.586,2305.028)
	(1498.633,2294.547)
	(1559.247,2284.753)
	(1617.518,2275.633)
	(1673.534,2267.171)
	(1727.386,2259.354)
	(1779.163,2252.170)
	(1828.953,2245.603)
	(1922.932,2234.268)
	(2010.037,2225.240)
	(2090.982,2218.409)
	(2166.481,2213.665)
	(2237.249,2210.899)
	(2304.000,2210.000)

\path(2304,2210)	(2370.751,2210.895)
	(2441.519,2213.659)
	(2517.018,2218.400)
	(2597.963,2225.229)
	(2685.068,2234.255)
	(2779.047,2245.589)
	(2828.837,2252.155)
	(2880.614,2259.340)
	(2934.466,2267.156)
	(2990.483,2275.617)
	(3048.753,2284.738)
	(3109.367,2294.532)
	(3172.414,2305.013)
	(3237.983,2316.194)
	(3306.162,2328.089)
	(3377.042,2340.713)
	(3450.712,2354.077)
	(3527.260,2368.197)
	(3606.777,2383.087)
	(3689.351,2398.759)
	(3775.072,2415.228)
	(3864.028,2432.506)
	(3956.310,2450.609)
	(4052.007,2469.550)
	(4151.207,2489.343)
	(4202.149,2499.562)
	(4254.000,2510.000)

\path(354,1310)	(405.851,1320.437)
	(456.793,1330.656)
	(555.993,1350.446)
	(651.690,1369.385)
	(743.972,1387.487)
	(832.928,1404.765)
	(918.649,1421.232)
	(1001.223,1436.903)
	(1080.740,1451.791)
	(1157.288,1465.911)
	(1230.958,1479.275)
	(1301.838,1491.897)
	(1370.017,1503.792)
	(1435.586,1514.972)
	(1498.633,1525.453)
	(1559.247,1535.247)
	(1617.518,1544.367)
	(1673.534,1552.829)
	(1727.386,1560.646)
	(1779.163,1567.830)
	(1828.953,1574.397)
	(1922.932,1585.732)
	(2010.037,1594.760)
	(2090.982,1601.591)
	(2166.481,1606.335)
	(2237.249,1609.101)
	(2304.000,1610.000)

\path(2304,1610)	(2370.751,1609.105)
	(2441.519,1606.341)
	(2517.018,1601.600)
	(2597.963,1594.771)
	(2685.068,1585.745)
	(2779.047,1574.411)
	(2828.837,1567.845)
	(2880.614,1560.660)
	(2934.466,1552.844)
	(2990.483,1544.383)
	(3048.753,1535.262)
	(3109.367,1525.468)
	(3172.414,1514.987)
	(3237.983,1503.806)
	(3306.162,1491.911)
	(3377.042,1479.287)
	(3450.712,1465.923)
	(3527.260,1451.803)
	(3606.777,1436.913)
	(3689.351,1421.241)
	(3775.072,1404.772)
	(3864.028,1387.494)
	(3956.310,1369.391)
	(4052.007,1350.450)
	(4151.207,1330.657)
	(4202.149,1320.438)
	(4254.000,1310.000)

\path(429,1385)	(394.528,1325.990)
	(361.812,1269.151)
	(330.839,1214.434)
	(301.595,1161.792)
	(248.240,1062.540)
	(201.636,971.010)
	(161.673,886.818)
	(128.242,809.578)
	(101.232,738.907)
	(80.535,674.420)
	(66.040,615.732)
	(57.636,562.459)
	(58.667,470.620)
	(82.748,395.825)
	(129.000,335.000)

\path(129,335)	(188.528,292.399)
	(263.212,262.497)
	(350.593,244.609)
	(448.216,238.047)
	(500.100,238.800)
	(553.622,242.126)
	(608.476,247.941)
	(664.354,256.159)
	(720.950,266.694)
	(777.955,279.461)
	(835.063,294.372)
	(891.968,311.344)
	(948.360,330.289)
	(1003.934,351.122)
	(1058.382,373.758)
	(1111.397,398.110)
	(1162.673,424.093)
	(1211.900,451.621)
	(1302.985,510.968)
	(1382.195,575.466)
	(1447.072,644.428)
	(1495.160,717.168)
	(1524.000,793.000)

\path(1524,793)	(1514.802,857.814)
	(1464.523,918.920)
	(1386.981,976.949)
	(1295.996,1032.531)
	(1205.387,1086.297)
	(1128.972,1138.877)
	(1080.570,1190.901)
	(1074.000,1243.000)

\path(1074,1243)	(1120.851,1306.713)
	(1204.990,1359.901)
	(1257.729,1381.767)
	(1315.813,1400.064)
	(1377.917,1414.480)
	(1442.715,1424.702)
	(1508.881,1430.418)
	(1575.090,1431.316)
	(1640.015,1427.081)
	(1702.332,1417.403)
	(1760.715,1401.968)
	(1813.837,1380.465)
	(1899.000,1318.000)

\path(1899,1318)	(1947.310,1218.057)
	(1948.978,1162.793)
	(1938.671,1104.582)
	(1918.649,1043.860)
	(1891.173,981.064)
	(1858.506,916.630)
	(1822.909,850.995)
	(1786.642,784.594)
	(1751.968,717.864)
	(1721.148,651.241)
	(1696.442,585.161)
	(1680.114,520.061)
	(1674.423,456.376)
	(1681.631,394.544)
	(1704.000,335.000)

\path(1704,335)	(1748.041,275.860)
	(1807.188,229.834)
	(1876.971,194.794)
	(1952.921,168.616)
	(2030.568,149.174)
	(2105.444,134.341)
	(2173.077,121.992)
	(2229.000,110.000)

\path(2229,110)	(2302.403,91.924)
	(2391.211,72.139)
	(2490.628,52.743)
	(2542.817,43.846)
	(2595.859,35.833)
	(2649.155,28.965)
	(2702.107,23.506)
	(2754.114,19.716)
	(2804.577,17.860)
	(2898.473,20.992)
	(2979.000,35.000)

\path(2979,35)	(3068.655,68.747)
	(3120.436,94.152)
	(3171.191,123.695)
	(3252.632,190.545)
	(3279.000,260.000)

\path(3279,260)	(3203.073,331.292)
	(3132.858,351.014)
	(3050.966,364.108)
	(2964.766,373.562)
	(2881.626,382.369)
	(2808.915,393.518)
	(2754.000,410.000)

\path(2754,410)	(2705.050,433.481)
	(2646.876,463.254)
	(2583.047,498.175)
	(2517.131,537.102)
	(2452.700,578.892)
	(2393.321,622.400)
	(2304.000,710.000)

\path(2304,710)	(2269.030,760.966)
	(2230.132,824.537)
	(2191.586,896.505)
	(2157.675,972.661)
	(2132.679,1048.798)
	(2120.881,1120.705)
	(2126.561,1184.175)
	(2154.000,1235.000)

\path(2154,1235)	(2231.989,1263.622)
	(2282.921,1258.608)
	(2337.548,1244.701)
	(2392.668,1224.894)
	(2445.082,1202.180)
	(2529.000,1160.000)

\path(2529,1160)	(2605.093,1100.750)
	(2686.384,1014.747)
	(2763.983,926.371)
	(2829.000,860.000)

\path(2829,860)	(2878.317,819.022)
	(2939.173,770.031)
	(3008.128,716.192)
	(3081.742,660.669)
	(3156.577,606.624)
	(3229.191,557.222)
	(3296.145,515.626)
	(3354.000,485.000)

\path(3354,485)	(3411.995,461.120)
	(3483.200,436.419)
	(3563.439,412.040)
	(3648.536,389.127)
	(3734.317,368.824)
	(3816.604,352.272)
	(3891.224,340.617)
	(3954.000,335.000)

\path(3954,335)	(4003.866,333.270)
	(4067.617,332.192)
	(4139.222,333.026)
	(4212.653,337.033)
	(4281.876,345.472)
	(4340.862,359.607)
	(4404.000,410.000)

\path(4404,410)	(4316.516,494.161)
	(4239.071,531.828)
	(4179.000,560.000)

\path(4179,560)	(4105.951,592.950)
	(4014.653,626.837)
	(3964.087,644.188)
	(3911.271,661.842)
	(3856.977,679.820)
	(3801.975,698.146)
	(3747.036,716.843)
	(3692.932,735.933)
	(3640.433,755.438)
	(3590.310,775.383)
	(3500.277,816.678)
	(3429.000,860.000)

\path(3429,860)	(3335.758,938.568)
	(3280.493,991.361)
	(3226.507,1047.980)
	(3179.084,1104.539)
	(3143.503,1157.152)
	(3129.000,1235.000)

\path(3129,1235)	(3196.049,1260.460)
	(3246.919,1249.156)
	(3307.411,1225.377)
	(3376.070,1191.780)
	(3451.441,1151.028)
	(3532.067,1105.778)
	(3616.492,1058.690)
	(3703.262,1012.425)
	(3790.920,969.641)
	(3878.011,932.999)
	(3963.078,905.158)
	(4044.666,888.778)
	(4121.320,886.519)
	(4191.583,901.040)
	(4254.000,935.000)

\path(4254,935)	(4304.445,996.608)
	(4321.260,1085.113)
	(4317.056,1142.747)
	(4304.445,1211.061)
	(4283.426,1291.372)
	(4254.000,1385.000)

\put(354,2735){\makebox(0,0)[b]{\large $i$}}
\put(4254,2735){\makebox(0,0)[b]{\large $j$}}
\put(354,1535){\makebox(0,0)[b]{\large $i$}}
\put(4254,1535){\makebox(0,0)[b]{\large $j$}}
\put(1554,1910){\makebox(0,0)[b]{\large $k$}}
\put(2004,635){\makebox(0,0)[b]{\large $k$}}
\end{picture}
}$$
\caption{\label{F-bridge-fig}Bridging and strong bridging sets.}
\end{figure}
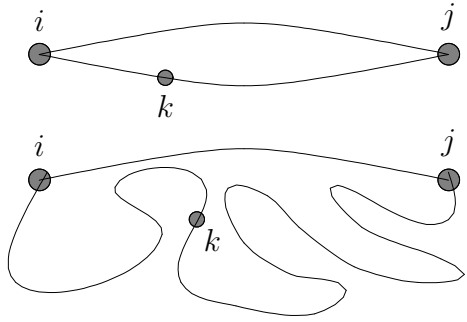

The difference between bridging sets and strong bridging sets is
depicted in Figure~\ref{F-bridge-fig}. All the paths shown there,
schematically, are shortest paths from~$i$ to~$j$ although they do not
all use the same number of edges. If $B$ is a strong $s$-bridging set,
and if $\eta(i,j)=t$ and $t\ge s$, i.e., if the minimum number of edges
on a shortest path from~$i$ to~$j$ is~$t$, and $t\ge s$, then there is a
vertex $k\in B$ that lies on a shortest path from~$i$ to~$j$ that uses
exactly~$t$ edges. The top drawing in Figure~\ref{F-bridge-fig}
illustrates the fact there there may be several shortest paths from~$i$
to~$j$ that use exactly~$t$ edges. A vertex~$k$ belonging to~$B$ is
guaranteed to lie on at least one of them. If $B$ is an $s$-bridging
set, but not necessarily a strong $s$-bridging set, then a vertex~$k$
belonging to~$B$ is guaranteed to lie on a shortest path from~$i$
to~$j$. But, this shortest path may use much more than~$t$ edges. This
is illustrated in the bottom drawing of Figure~\ref{F-bridge-fig}.

It is not difficult to see that if~$s$ is an integer then we can replace
the condition $\eta(i,j)\ge s$ in the definition of bridging, and
strongly bridging, sets by the condition $\eta(i,j)=s$. Indeed, suppose
the appropriate condition holds for every $u,v\in V$ such that
$\eta(u,v)=s$. Suppose that $\eta(i,j)=t>s$. Consider a shortest
path~$p$ from~$i$ to~$j$ that uses~$t$ edges. Let~$w$ be the $s$-th
vertex on~$p$, starting the count from~$0$. Then, clearly $\eta(i,w)=s$.
Thus, a vertex $k\in B$ is guaranteed to lie on a shortest path from~$i$
to~$w$. This vertex lies also on a shortest path from~$i$ to~$j$, or on
such a shortest path with a minimum number of edges, as required.

Reviewing the proof of Lemma~\ref{L-rand}, we see that algorithm
\RANDSHORTPATH\ produces correct results as long as the set~$B$ used in the
$\ell$-th iteration is a {\em strong\/} $(s/3)$-bridging set.

\begin{lemma}\label{L-bridge}
If in each iteration of \RANDSHORTPATH\ the set~$B$ is a strong
$(s/3)$-bridging set, then all distances returned by \RANDSHORTPATH\ are
correct. 
\end{lemma}

\begin{proof}
    The proof is almost identical to the proof of Lemma~\ref{L-rand}. We
    show again, by induction on~$\ell$, that if $\eta(i,j)\le
    (3/2)^\ell$, then after the $\ell$-th iteration of the algorithm we
    have $f_{ij}=\del(i,j)$. The basis of the induction is easily
    established. Suppose, therefore, that the claim holds for $\ell-1$.
    We show that it also holds for~$\ell$. Let~$i$ and~$j$ be two
    vertices such that $2s/3 \le \eta(i,j)\le s$, where
    $s=(2/3)^{\ell}$. As in Lemma~\ref{L-rand}, let~$p$ be a shortest
    path from~$i$ to~$j$ that uses $\eta(i,j)$ edges, let~$I$ and~$J$ be
    two vertices on~$p$ such that~$I$ and~$J$ are separated, on~$p$, by
    $s/3$ edges, and such that~$i$ and~$I$, and~$J$ and~$j$, are
    separated, on~$p$, by at most $s/3$ edges (see Figure~\ref{F-cor}).
    As~$B$, the set used in the $\ell$-th iteration, is assumed to be a
    strong $(s/3)$-bridging set, and as $\eta(I,J)\ge s/3$, a vertex
    $k\in B$ is guaranteed to lie of a shortest path from~$I$ to~$J$
    that uses $\eta(I,J)$ edges.  This shortest path from~$I$ to~$J$ is
    not necessarily the portion of~$p$ going from~$I$ to~$J$.
    Nonetheless, we still have $\del(i,j)=\del(i,k)+\del(k,j)$ and
    $\eta(i,j)=\eta(i,k)+\eta(k,j)$.  As $\eta(i,k)\le \eta(i,J)\le
    2s/3$ and $\eta(k,j)\le \eta(I,j)\le 2s/3$ we get, by the induction
    hypothesis, that $f_{ik}=\del(i,k)$ and $f_{kj}=\del(k,j)$. After
    the distance product of the $\ell$-th iteration we therefore have
    $f_{ij}=\del(i,j)$, as required.
\end{proof}

In the proof of Lemma~\ref{L-bridge} we made heavy use of the
assumption that~$B$ is a {\em strong\/} bridging set. If~$B$ were not a
strong bridging set, we could not have deduced that
$\eta(i,k),\eta(k,j)\le 2s/3$ and the argument used in the proof would
break down.
Also implicit in the proof of Lemma~\ref{L-rand} is the following result
whose proof we do not repeat:

\begin{lemma}
    Let $G=(V,E)$ be a weighted directed graph on~$n$ vertices and let
    $s\ge 1$. If~$B$ is a random set obtained by running
    $\RAND(\{1,2,\ldots,n\},(3\ln n)/s)$, i.e., if each vertex
    of~$V$ is added to~$B$ independently with probability $(3\ln
    n)/s$, then with very high probability~$B$ is a
    {\em strong\/} $s$-bridging set.
\end{lemma}

\begin{figure}[t]
\begin{center}
    \framebox{\hspace{0.6cm}\parbox{4.8in}{
\algorithm\ \FINDBRIDGE$(W,s)$ $\vphantom{2^{2^{2^2}}}$ \\[5pt]
$\CC\gets \phi$\\[3pt]
{\tt for every} $1\le i,j\le n$ {\tt do}\\[3pt]
$\null\qquad U\gets \SUBPATH(W,i,j,s-1)\,\cup\,\{i,j\}$\\
$\null\qquad${\tt if} $|U|\ge s$ {\tt then} 
$\CC\gets\CC\cup \{U\}$ {\tt endif}\\[3pt]
{\tt end}\\[3pt]
$B\gets\HITTINGSET(\CC)$\\[3pt]
{\tt return} $B$ $\phantom{2_{2_{2_2}}}$ }\hspace{0.6cm}}
 \end{center}

\caption{\label{F-bridge}A deterministic algorithm for constructing an
     $s$-bridging set.}
\end{figure}

\begin{figure}[t]
\begin{center}
\null\quad\\
\framebox{\hspace{0.4cm}\parbox{4.8in}{
\algorithm\ \SUBPATH$(W,i,j,s)$ $\phantom{2^{2^{2^2}}}$ \\[5pt]
{\tt if} $w_{ij}=0$ {\tt or} $s=0$ {\tt then} \\
$\null\qquad${\tt return} $\phi$\\
{\tt else} \\
$\null\qquad U\gets \SUBPATH(W,i,w_{ij},s-1)$\\
$\null\qquad${\tt return} 
$U\,\cup\,\{w_{ij}\}\,\cup\,\SUBPATH(W,w_{ij},j,s-|U|-1)$\\
{\tt endif}$\phantom{2_{2_{2_2}}}$
}\hspace{0.6cm}}
 \end{center}
\caption{\label{F-subpath}Finding up to $s$ vertices on a 
   shortest path from~$i$ to~$j$.}
\end{figure}

We next describe a deterministic algorithm, called \FINDBRIDGE, for
finding $s$-bridging sets. (Unfortunately, the sets returned by
\FINDBRIDGE\ are not necessarily strong $s$-bridging sets.) A
description of algorithm \FINDBRIDGE\ is given in Figure~\ref{F-bridge}.
It receives an $n\times n$ matrix of witnesses~$W$ using which it is
possible to reconstruct shortest paths between all pairs of vertices
$i,j\in V$ for which $\eta(i,j)\le s$. In other words, if $\eta(i,j)\le
s$, then $\PATH(W,i,j)$ produces a shortest path from~$i$ to~$j$. We
assume here, for simplicity, that the graph does not contain cycles of
non-positive weight so the shortest path produced by $\PATH(W,i,j)$,
when $\eta(i,j)\le s$, is simple. We show later how to remove this
simplifying assumption. We do {\em not\/} assume that the shortest path
produced by $\PATH(W,i,j)$ uses a minimum number of edges, i.e., it may
use more than $\eta(i,j)$ edges.

Algorithm \FINDBRIDGE\ uses a procedure called \SUBPATH\ that receives
the matrix~$W$, two vertices $i,j\in V$ and an integer~$s$. The
operation of \SUBPATH\ is similar to the operation of \PATH. It tries to
construct a path from~$i$ to~$j$ using the witnesses contained in the
matrix~$W$. It counts, however, the number of intermediate vertices
found so far on the path and stops the construction when~$s$
intermediate vertices are encountered. A simple recursive implementation
of \SUBPATH\ is given in Figure~\ref{F-subpath}. The following lemma is
easily verified.

\begin{lemma}
    If a call to $\PATH(W,i,j)$ constructs a simple path from $i$ to $j$
    that passes through $t$ intermediate vertices, then
    $\SUBPATH(W,i,j,s)$ returns the set of intermediate vertices on this
    path, if $t\le s$, or a subset of $s$ intermediate vertices on this
    path, if $t>s$. The running time of $\SUBPATH(W,i,j,s)$ is $O(s)$.
\end{lemma}

For every $i,j\in V$, let $U_{ij}$ be the set obtained by adding the
vertices $i$ and~$j$ to the set obtained by calling $\SUBPATH(W,i,j,s)$.
All the elements of~$U_{ij}$ are vertices on a shortest path from~$i$
to~$j$. If $\eta(i,j)=s$, then by our assumption on~$W$, $\PATH(W,i,j)$
returns a shortest path from~$i$ to~$j$. This shortest path must
use at least~$s$ edges and contain, therefore, at least $s-1$
intermediate vertices. It follows that $|U_{ij}|\ge s+1$. Thus, if a
set $B$ {\em hits\/} all the sets~$U_{ij}$ for which $|U_{ij}|\ge s$,
i.e., if $B\cap U_{ij}\ne\phi$ whenever $|U_{ij}|\ge s$, then $B$ is
$s$-bridging. Algorithm \FINDBRIDGE\ collects all the sets~$U_{ij}$ for
which $|U_{ij}|\ge s$ into a collection of sets called~$\CC$. It then
calls algorithm \HITTINGSET\ to find a set that hits all the sets in
this collection.

Algorithm \HITTINGSET\ uses the greedy heuristic to find a set~$B$ that
hits all the sets in the collection~$\CC$. As shown by Lov\'{a}sz
\cite{Lovasz75} and Chv\'{a}tal \cite{Chvatal79}, the size of the
hitting set returned by \HITTINGSET\ is at most $(\ln \Delta)+1$ times
the size of the optimal {\em fractional\/} hitting set, where~$\Delta$
is the maximal number of sets that a single element can hit. As each set
in the collection~$\CC$ contains at least~$s$ elements, there is a fractional
hitting set of size~$n/s$. This fractional hitting set is obtained by
giving each one of the~$n$ vertices of $V$ a weight of $1/s$. As there
are at most $n^2$ sets to hit, we get that $\Delta\le n^2$. As a
consequence we get that \FINDBRIDGE\ returns a bridging set of size at
most $n(2\ln n+1)/s$. \HITTINGSET\ can be easily implemented to run in
time which is linear in the sum of the sizes of the sets in the
collection.  The running time of \FINDBRIDGE\ is therefore easily seen
to be $O(n^2s)$. We obtained, therefore, the following result:

\ignore{
It is also not difficult to see that if the shortest paths encoded by~$W$
use a minimum number of edges, then the set~$B$ obtained by running
\FINDBRIDGE\ is a strong $s$-bridging set. (We shall see, however, that
it is not so easy to satisfy this condition.) 
We obtained, therefore, the following result:
}

\begin{lemma}
  If the matrix~$W$ can be used to construct shortest paths between
  all pairs of vertices $i,j\in V$ for which $\eta(i,j)\le s$, then
  algorithm \FINDBRIDGE\ finds an $s$-bridging set of size at most
  $n(2\ln n+1)/s$. The running time of \FINDBRIDGE\ is $O(n^2s)$.
\end{lemma}

Unfortunately, the sets returned by
\FINDBRIDGE\ are not necessarily strong bridging sets. But,
if the input graph is {\em unweighted}, then an $s$-bridging set is
also a strong $s$-bridging set. Thus, if we replace the call to
\RAND\ in \RANDSHORTPATH\ by
$$\begin{array}{l}
\mbox{{\tt if} $s\le n^{1/2}$ {\tt then} }\\
\mbox{$\qquad B\gets \FINDBRIDGE(W,\lfloor s/3 \rfloor)$}\\
\mbox{{\tt endif}}\\
\end{array}$$
we obtain a deterministic algorithm for solving the APSP problem for
{\em unweighted\/} directed graphs. We call this algorithm
\UNWEIGHTEDSHORTPATH.

We compute new bridging sets only when $s\le n^{1/2}$ as computing
bridging sets for larger values of $s$ may consume too much
time. (Recall that the running time of \FINDBRIDGE\ is $O(n^2s)$.) The
algorithm remains correct as an $s$-bridging set is also an
$s'$-bridging set for every $s'\ge s$. The use of a bridging set of
size $\Theta(n^{1/2}\log n)$ in the iterations for which $s\ge
n^{1/2}$ does not change the overall running time of the algorithm, as
in all these iterations the required distance product can be computed
using the naive algorithm in $\Ot(n^{2.5})$ time. We thus get:

\begin{theorem}
  Algorithm \UNWEIGHTEDSHORTPATH\ solves the APSP problem for unweighted
  directed graphs {\em deterministically\/} in $\Ot(n^{2+\mu})$ time,
  where $\mu<0.575$ satisfies $\omega(1,\mu,1)=1+2\mu$.
\end{theorem}

\section{A deterministic algorithm for weighted graphs}
\label{S-weighted}

In this section we present a deterministic version of algorithm
\RANDSHORTPATH\ for weighted directed graphs. The algorithm, called
\SHORTPATH\ is given in Figure~\ref{F-SHORTPATH}. For simplicity, we
assume at first that the input graph does not contain negative weight
cycles, nor zero weight cycles. 

\begin{figure}[t]
\begin{center}
\framebox{\hspace{0.6cm}\parbox{4.8in}{
{\tt algorithm} $\SHORTPATH(D) \phantom{2^{2^{2^2}}}$ \\[5pt]
$F\gets D$ ; $W\gets 0$ \\
$M\gets \max\{\;|d_{ij}| : d_{ij}\ne+\infty\}$\\[3pt]
{\tt for} $\ell\gets 1$ {\tt to} $\lceil\log_{2}n\rceil$ {\tt do}\\
{\tt begin}\\[3pt]
$\null\qquad s\gets 2^\ell$\\[3pt]
$\null\qquad${\tt if} $s\le n^{1/2}$ {\tt then}\\
$\null\qquad\qquad B\gets \FINDBRIDGEUPD(F,W,s/2)$\\
$\null\qquad${\tt endif}\\[5pt]
%
%
$\null\qquad \DISTPRODUPD(F,W,\,\,B,V,V,\,\phantom{2}sM) $\\
$\null\qquad \DISTPRODUPD(F,W,\,\,V,B,V,\,2sM) $\\[3pt]
{\tt end}\\[3pt]
{\tt return} $(F,W)$ $\phantom{2_{2_{2_2}}}$
}\hspace{0.6cm}}
\end{center}
\caption{\label{F-SHORTPATH}A deterministic algorithm for finding
   shortest paths.}

\vspace*{10pt}
\begin{center}
\framebox{\hspace{0.6cm}\parbox{4.8in}{
{\tt algorithm} $\DISTPRODUPD(F,W,\,A,B,C,\,L) \phantom{2^{2^{2^2}}}$ \\[5pt]
$(F',W')\gets \DISTPROD(F[A,B],F[B,C],L) $\\[3pt]
{\tt for every} $1\le i\le |A|$ {\tt and} $1\le j\le |C|$ {\tt do}\\
{\tt if} $f'_{ij}<f_{a_i,c_j}$ {\tt then}
$f_{a_i,c_j}\gets f'_{ij}$ ; $w_{a_i,c_j}\gets b_{w'_{ij}}$ {\tt
   endif}$\phantom{2_{2_{2_2}}}$ 
}\hspace{0.6cm}}
\end{center}
\caption{\label{F-DISTPRODUPD}A simple procedure for updating
   distances and witnesses.}
\end{figure}

\begin{figure}[t]
\begin{center}
    \framebox{\hspace{0.6cm}\parbox{4.8in}{
\algorithm\ \FINDBRIDGEUPD$(F,W,s)$ $\vphantom{2^{2^{2^2}}}$ \\[5pt]
$\CC\gets \phi$\\[3pt]
{\tt for every} $1\le i,j\le n$ {\tt do}\\[3pt]
$\null\qquad U\gets \SUBPATHUPD(F,W,i,j,i,j,s-1)\,\cup\,\{i,j\}$\\
$\null\qquad${\tt if} $|U|\ge s$ {\tt then} 
$\CC\gets\CC\cup \{U\}$ {\tt endif}\\[3pt]
{\tt end}\\[3pt]
$B\gets\HITTINGSET(\CC)$\\[3pt]
{\tt return} $B$ $\phantom{2_{2_{2_2}}}$ }\hspace{0.6cm}}
 \end{center}

\caption{\label{F-FINDBRIDGEUPD}A deterministic algorithm for constructing an
     $s$-bridging set while updating some distances.}

\vspace*{10pt}
\begin{center}
\framebox{\hspace{0.4cm}\parbox{4.8in}{
\algorithm\ \SUBPATHUPD$(F,W,a,b,i,j,s)$ $\phantom{2^{2^{2^2}}}$ \\[5pt]
{\tt if} $w_{ij}=0$ {\tt or} $s=0$ {\tt then} \\
$\null\qquad${\tt return} $\phi$\\
{\tt else} \\
\null\qquad{\tt if} $f_{ai}+f_{i,w_{ij}}< f_{a,w_{ij}}$ {\tt then}
$f_{a,w_{ij}} \gets f_{ai}+f_{i,w_{ij}}$ ; $w_{a,w_{ij}}\gets i\;$ 
{\tt endif}\\
\null\qquad{\tt if} $f_{w_{ij},j}+f_{j,b}< f_{w_{ij},b}$ {\tt then}
$f_{w_{ij},b} \gets f_{w_{ij},j}+f_{j,b}$ ; $f_{w_{ij},b}\gets j\;$
{\tt endif}\\[5pt]
\null\qquad$U\gets \SUBPATHUPD(F,W,a,b,i,w_{ij},s-1)$\\
$\null\qquad${\tt return} 
$U\,\cup\,\{w_{ij}\}\,\cup\,\SUBPATHUPD(F,W,a,b,w_{ij},j,s-|U|-1)$\\
{\tt endif}$\phantom{2_{2_{2_2}}}$
}\hspace{0.6cm}}
 \end{center}
\caption{\label{F-SUBPATHUPD}Finding up to $s$ vertices on a 
   shortest path from~$i$ to~$j$ while updating distances.}
\ignore{
\vspace*{10pt}
\begin{center}
\framebox{\hspace{0.4cm}\parbox{4.8in}{
\algorithm\ \UPDATE$(f,w,f',w')$ $\phantom{2^{2^{2^2}}}$ \\[5pt]
{\tt if} $f'<f$ {\tt then} $f\gets f'$ ; $w\gets w'$ {\tt
   endif}$\phantom{2_{2_{2_2}}}$ 
}\hspace{0.6cm}}
 \end{center}
\caption{\label{F-update}A simple procedure for updating a single
   distance and its corresponding witness.}
}
\end{figure}

Algorithm \SHORTPATH\ uses a simple procedure, called \DISTPRODUPD, that
performs a distance product, by calling \DISTPROD\ of
Section~\ref{S-prod}, and updates the distances and witnesses found so
far. Algorithm \DISTPRODUPD\ is given in Figure~\ref{F-DISTPRODUPD}.  It
receives the $n\times n$ matrices~$F$ and~$W$ that hold the distances
and witnesses found so far. It also receives three subsets
$A,B,C\subseteq V$, where $V=\{1,2,\ldots,n\}$ is the set of all
vertices. (In the calls made by \SHORTPATH, two of the sets $A,B$ and
$C$ would be~$V$.)  \DISTPRODUPD\ computes the distance product
$F[A,B]\star F[B,C]$, putting a cap of~$L$ of the values of the entries
of~$F$ that participate in the product. It then updates the matrices~$F$
and~$W$ accordingly. (By $F[A,B]$ we obviously mean the submatrix of~$F$
composed of the elements whose row index belongs to~$A$, and whose
column index belongs to~$B$. Also, we let $a_i$ denote the $i$-th
elements of~$A$.) Thus, the first call to \DISTPRODUPD\ in \SHORTPATH\ 
computes the distance product $F[B,*]\star F$, while the second one
computes the distance product $F[*,B]\star F[B,*]$, as in \RANDSHORTPATH. By
Lemma~\ref{L-11r}, we get that the cost of these two distance products
is essentially the same.

Algorithm \SHORTPATH\ constructs bridging sets by calling algorithm
\FINDBRIDGEUPD\ given in Figure~\ref{F-FINDBRIDGEUPD}. Algorithm
\FINDBRIDGEUPD\ is very similar to algorithm \FINDBRIDGE\ of
Section~\ref{S-deter}. The difference is that \FINDBRIDGEUPD\ calls
algorithm \SUBPATHUPD, given in Figure~\ref{F-SUBPATHUPD}, instead of
algorithm \SUBPATH\ called by \FINDBRIDGE.

A call to $\SUBPATH(W,i,j,s)$ returns a set of up to~$s$ intermediate
vertices on a path from~$i$ to~$j$. However, if $k\in
\SUBPATH(W,i,j,s)$, it is not guaranteed that $f_{ik},f_{kj}<+\infty$,
let alone $f_{ik}+f_{kj}\le f_{ij}$. Algorithm \SUBPATHUPD\ fixes this
problem. The following lemma is easily verified.
\ignore{ If $k\in \SUBPATH(F,W,i,j,i,j,s)$, then $f_{ik}+f_{kj}\le
f_{ij}$ while still having $f_{ik}\ge\del(i,k)$ and $f_{kj}\ge
\del(k,j)$. Thus in particular, if $f_{ij}=\del(i,j)$, then after the
call to \SUBPATHUPD\ we have $f_{ik}=\del(i,k)$, $f_{kj}=\del(k,j)$ and
$\del(i,j)=\del(i,k)+\del(k,j)$.}

\begin{lemma}\label{L-SUBPATHUPD}
    If the matrices~$F$ and~$W$ satisfies the conditions $f_{ij}\ge
    \del(i,j)$, for every $i,j\in V$, and $f_{ij}\ge
    f_{i,w_{ij}}+f_{w_{ij},j}$ whenever $1\le w_{ij}\le n$, and if
    $\PATH(W,i,j)$ traces a path from~$i$ to~$j$, then a call to
    $\SUBPATHUPD(F,W,i,j,i,j,s)$ returns a set of~$s$ intermediate
    vertices on this path, or the set of all intermediate vertices if
    there are less than~$s$ of them. If~$k$ is one of the vertices
    returned by the call, then after the call we have $f_{ik}+f_{kj}\le
    f_{ij}$. The matrices~$F$ and~$W$ continue to satisfy the specified
    conditions.  Furthermore, if before the call we have
    $f_{ij}=\del(i,j)$, then after the call to we have
    $f_{ik}=\del(i,k)$, $f_{kj}=\del(k,j)$ and
    $\del(i,j)=\del(i,k)+\del(k,j)$.
\end{lemma}

Before proving the correctness of algorithm \SHORTPATH, we prove a useful
additional property of bridging sets.

\begin{lemma}\label{L-close}
Let $B$ be an $s$-bridging set of a graph $G=(V,E)$ with no nonpositive weight
cycles. Then, if $i,j\in V$ and $\eta(i,j)\ge s$, then there is a vertex
$k\in B$ such that $\del(i,j)=\del(i,k)+\del(k,j)$ and $\eta(i,k)\le s$.
\end{lemma}

\begin{proof}
    By the definition of bridging sets, we get that there exists $k_1\in
    B$ such that $\del(i,j)=\del(i,k_1)+\del(k_1,j)$. If $\eta(i,k_1)\le
    s$, we are done. Assume, therefore, that $\eta(i,k_1)>s$. Let $k'_1$
    be next to last vertex on a shortest path from~$i$ to~$k_1$.
    Clearly, $k'_1\ne k_1$, $\del(i,j)=\del(i,k'_1)+\del(k'_1,j)$ and
    $\eta(i,k'_1)\ge s$. Thus, there exists $k_2\in B$ such that
    $\del(i,k'_1)=\del(i,k_2)+\del(k_2,k'_1)$, and therefore also
    $\del(i,j)=\del(i,k_2)+\del(k_2,j)$. There is, therefore, a shortest
    path from~$i$ to~$j$ that passes through $k_2$, then through~$k'_1$,
    and then through~$k_1$. As there are no nonpositive weight cycles in
    the graph, a shortest path must be simple and therefore $k_2\ne
    k_1$. In general, suppose that we have found so far~$r$ distinct
    vertices $k_r,k_{r-1},\ldots,k_1\in B$ such that there is a shortest
    path from~$i$ to~$j$ that visits all these vertices.
    If~$\eta(i,k_r)\le s$, then we are done. Otherwise, we can find
    another vertex $k_{r+1}\in B$, distinct from all the previous
    vertices, such that there is a shortest path from~$i$ to~$j$ that
    passes though $k_{r+1},k_r,k_{r-1},\ldots,k_1$. As the graph is
    finite, this process must eventually end with a vertex from~$B$
    satisfying our requirements.
\end{proof}

\begin{theorem}
    Algorithm \SHORTPATH\ finds all
    distances, and a succinct representation of
    shortest paths between all pairs of vertices in a graph with no
    nonpositive weight cycles. If the 
    input graph has~$n$ vertices and the edge
    weights are taken from the set $\{-M,\ldots,0,\ldots,M\}$, where 
    $M=n^t$ and $t\le 3-\omega$, then its
    running time is $\Ot(n^{2+\mu(t)})$, where $\mu=\mu(t)$ satisfies
    $\omega(1,\mu,1)=1+2\mu-t$.
\end{theorem}

\begin{proof}
We prove, by induction, that after the $\ell$-th iteration of
\SHORTPATH\ we have:
\begin{enumerate}
\item[(i)] $f_{ij}\ge \del(i,j)$, for every $i,j\in V$.
\item[(ii)] If $w_{ij}=0$ then $f_{ij}=d_{ij}$. Otherwise, $1\le w_{ij}\le n$
and $f_{ij}\ge f_{i,w_{ij}}+f_{w_{ij},j}$.
\item[(iii)] If $\eta(i,j)\le 2^\ell$, then $f_{ij}=\del(i,j)$.
\end{enumerate}

The proofs of properties~(i) and~(ii) are analogous to the proofs of
properties~(i) and~(ii) of Lemma~\ref{L-simple}. We concentrate,
therefore, on the proof of property~(iii). It is easy to check that
property~(iii) holds before the first iteration. We show now that if it
holds at the end of the $(\ell-1)$-st iteration, then it also holds
after the $\ell$-th iteration.

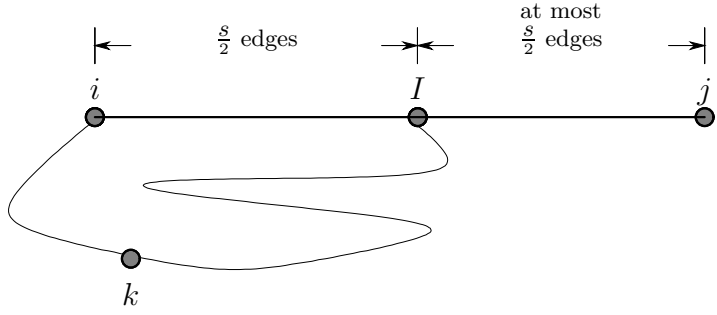
\begin{figure*}[t]
\newcommand{\ls}{\mbox{$s$}}
\newcommand{\AAA}{\shortstack[c]{at most \\[0pt] $\frac{\ls}{2}$ edges}}
\newcommand{\BBB}{$\frac{\ls}{2}$ edges}
\newcommand{\CCC}{\AAA}
$$\setlength{\unitlength}{0.00062500in}
\begingroup\makeatletter\ifx\SetFigFont\undefined%
\gdef\SetFigFont#1#2#3#4#5{%
  \reset@font\fontsize{#1}{#2pt}%
  \fontfamily{#3}\fontseries{#4}\fontshape{#5}%
  \selectfont}%
\fi\endgroup%
{\renewcommand{\dashlinestretch}{30}
\begin{picture}(6187,2410)(0,-10)
\path(704,1588)(703,1587)(699,1584)
	(693,1579)(684,1571)(671,1560)
	(653,1546)(632,1528)(606,1506)
	(576,1481)(542,1452)(505,1420)
	(465,1385)(424,1349)(381,1311)
	(338,1271)(296,1232)(254,1192)
	(214,1153)(177,1114)(142,1077)
	(111,1041)(83,1007)(60,975)
	(41,944)(26,916)(17,889)
	(12,864)(12,840)(18,818)
	(28,798)(44,778)(59,764)
	(77,750)(98,736)(121,722)
	(148,709)(176,696)(208,683)
	(242,671)(278,658)(316,646)
	(356,633)(399,621)(443,609)
	(489,597)(536,584)(585,572)
	(634,561)(685,549)(737,537)
	(789,525)(842,514)(895,503)
	(948,491)(1001,481)(1055,470)
	(1107,459)(1160,449)(1212,440)
	(1263,430)(1314,421)(1364,413)
	(1413,405)(1462,397)(1509,390)
	(1555,384)(1601,378)(1646,373)
	(1690,368)(1733,364)(1775,361)
	(1817,359)(1859,358)(1903,358)
	(1947,358)(1991,359)(2036,362)
	(2082,365)(2129,369)(2177,374)
	(2227,380)(2278,386)(2330,394)
	(2383,402)(2438,410)(2493,419)
	(2550,429)(2607,439)(2665,449)
	(2723,460)(2781,471)(2839,482)
	(2897,494)(2954,506)(3010,517)
	(3065,529)(3118,541)(3169,552)
	(3219,564)(3266,575)(3310,586)
	(3351,597)(3389,608)(3424,618)
	(3454,629)(3481,639)(3503,648)
	(3521,658)(3535,667)(3543,676)
	(3547,685)(3545,694)(3539,703)
	(3530,710)(3517,718)(3500,725)
	(3478,733)(3453,740)(3423,748)
	(3390,755)(3352,763)(3309,771)
	(3263,779)(3213,787)(3159,794)
	(3101,803)(3040,811)(2976,819)
	(2909,827)(2839,835)(2767,843)
	(2693,852)(2617,860)(2540,868)
	(2462,877)(2383,885)(2305,894)
	(2226,902)(2147,910)(2070,918)
	(1993,927)(1918,935)(1845,943)
	(1774,951)(1706,958)(1640,966)
	(1578,974)(1519,981)(1463,989)
	(1412,996)(1364,1003)(1321,1010)
	(1283,1016)(1248,1023)(1219,1029)
	(1194,1035)(1174,1041)(1158,1047)
	(1147,1053)(1141,1058)(1139,1063)
	(1141,1068)(1148,1073)(1159,1077)
	(1174,1081)(1194,1085)(1218,1089)
	(1247,1092)(1280,1095)(1317,1098)
	(1358,1101)(1403,1104)(1452,1106)
	(1504,1108)(1560,1110)(1618,1111)
	(1680,1113)(1744,1114)(1810,1115)
	(1879,1117)(1949,1118)(2020,1118)
	(2093,1119)(2167,1120)(2241,1121)
	(2315,1122)(2390,1123)(2464,1123)
	(2537,1124)(2610,1125)(2681,1126)
	(2751,1128)(2819,1129)(2885,1131)
	(2949,1133)(3011,1134)(3070,1137)
	(3126,1139)(3180,1142)(3231,1145)
	(3279,1148)(3324,1151)(3365,1155)
	(3404,1159)(3439,1163)(3472,1168)
	(3502,1173)(3529,1178)(3554,1183)
	(3591,1193)(3622,1203)(3646,1215)
	(3664,1228)(3676,1243)(3683,1258)
	(3685,1275)(3682,1294)(3674,1314)
	(3662,1334)(3647,1356)(3630,1379)
	(3609,1402)(3587,1425)(3564,1448)
	(3541,1470)(3518,1491)(3496,1511)
	(3476,1528)(3458,1544)(3442,1557)
	(3429,1567)(3419,1576)(3412,1581)
	(3408,1585)(3405,1587)(3404,1588)
\texture{aaffffff ffaaaaaa aaffffff ffaaaaaa aaffffff ffaaaaaa aaffffff ffaaaaaa 
	aaffffff ffaaaaaa aaffffff ffaaaaaa aaffffff ffaaaaaa aaffffff ffaaaaaa 
	aaffffff ffaaaaaa aaffffff ffaaaaaa aaffffff ffaaaaaa aaffffff ffaaaaaa 
	aaffffff ffaaaaaa aaffffff ffaaaaaa aaffffff ffaaaaaa aaffffff ffaaaaaa }
\path(5834,2383)(5834,2083)
\path(5834,2383)(5834,2083)
\path(5534,2233)(5834,2233)
\path(5534,2233)(5834,2233)
\blacken\path(5714.000,2203.000)(5834.000,2233.000)(5714.000,2263.000)(5750.000,2233.000)(5714.000,2203.000)
\thicklines
\texture{44555555 55aaaaaa aa555555 55aaaaaa aa555555 55aaaaaa aa555555 55aaaaaa 
	aa555555 55aaaaaa aa555555 55aaaaaa aa555555 55aaaaaa aa555555 55aaaaaa 
	aa555555 55aaaaaa aa555555 55aaaaaa aa555555 55aaaaaa aa555555 55aaaaaa 
	aa555555 55aaaaaa aa555555 55aaaaaa aa555555 55aaaaaa aa555555 55aaaaaa }
\put(3434,1633){\shade\ellipse{150}{150}}
\put(3434,1633){\ellipse{150}{150}}
\put(734,1633){\shade\ellipse{150}{150}}
\put(734,1633){\ellipse{150}{150}}
\put(3434,1633){\shade\ellipse{150}{150}}
\put(3434,1633){\ellipse{150}{150}}
\put(734,1633){\shade\ellipse{150}{150}}
\put(734,1633){\ellipse{150}{150}}
\put(5834,1633){\shade\ellipse{150}{150}}
\put(5834,1633){\ellipse{150}{150}}
\put(1034,448){\shade\ellipse{150}{150}}
\put(1034,448){\ellipse{150}{150}}
\thinlines
\texture{aaffffff ffaaaaaa aaffffff ffaaaaaa aaffffff ffaaaaaa aaffffff ffaaaaaa 
	aaffffff ffaaaaaa aaffffff ffaaaaaa aaffffff ffaaaaaa aaffffff ffaaaaaa 
	aaffffff ffaaaaaa aaffffff ffaaaaaa aaffffff ffaaaaaa aaffffff ffaaaaaa 
	aaffffff ffaaaaaa aaffffff ffaaaaaa aaffffff ffaaaaaa aaffffff ffaaaaaa }
\path(734,2383)(734,2083)
\path(734,2383)(734,2083)
\blacken\path(854.000,2263.000)(734.000,2233.000)(854.000,2203.000)(818.000,2233.000)(854.000,2263.000)
\path(734,2233)(1034,2233)
\path(734,2233)(1034,2233)
\path(3434,2383)(3434,2083)
\path(3434,2383)(3434,2083)
\path(3134,2233)(3434,2233)
\path(3134,2233)(3434,2233)
\blacken\path(3314.000,2203.000)(3434.000,2233.000)(3314.000,2263.000)(3350.000,2233.000)(3314.000,2203.000)
\blacken\path(3554.000,2263.000)(3434.000,2233.000)(3554.000,2203.000)(3518.000,2233.000)(3554.000,2263.000)
\path(3434,2233)(3734,2233)
\path(3434,2233)(3734,2233)
\thicklines
\path(734,1633)(5834,1633)
\put(2084,2233){\makebox(0,0)[b]{\smash{{{\SetFigFont{9}{10.8}{\rmdefault}{\mddefault}{\updefault}\BBB}}}}}
\put(4634,2233){\makebox(0,0)[b]{\smash{{{\SetFigFont{9}{10.8}{\rmdefault}{\mddefault}{\updefault}\CCC}}}}}
\put(1034,58){\makebox(0,0)[b]{\smash{{{\SetFigFont{9}{10.8}{\rmdefault}{\mddefault}{\updefault}\large $k$}}}}}
\put(734,1783){\makebox(0,0)[b]{\smash{{{\SetFigFont{9}{10.8}{\rmdefault}{\mddefault}{\updefault}\large $i$}}}}}
\put(3434,1783){\makebox(0,0)[b]{\smash{{{\SetFigFont{9}{10.8}{\rmdefault}{\mddefault}{\updefault}\large $I$}}}}}
\put(5834,1783){\makebox(0,0)[b]{\smash{{{\SetFigFont{9}{10.8}{\rmdefault}{\mddefault}{\updefault}\large $j$}}}}}
\end{picture}
}$$ 
\caption{\label{F-proof}The correctness proof of \SHORTPATH.}
\end{figure*}

Let $i,j\in V$ be such that $\eta(i,j)\le 2^\ell$. If $\eta(i,j)\le
2^{\ell-1}$, then the condition $f_{ij}=\del(i,j)$ holds already after
the $(\ell-1)$-st iteration. Assume, therefore, that
$2^{\ell-1}<\eta(i,j)\le 2^\ell$. Let~$p$ be a shortest path from~$i$
to~$j$ that uses $\eta(i,j)$ edges. Let $I$ be the vertex on~$p$ for
which $\eta(i,I)=2^{\ell-1}$. (See Figure~\ref{F-proof}.) Note that
$\eta(I,j)\le 2^{\ell-1}$. By the induction hypothesis, after
the $(\ell-1)$-st iteration we have $f_{iI}=\del(i,I)$ and
$f_{Ij}=\del(I,j)$.  As~$B$ is an $2^{\ell-1}$-bridging set, we get, by
Lemma~\ref{L-close}, that there exists $k\in B$ such that
$\del(i,I)=\del(i,k)+\del(k,I)$ and $\eta(i,k)\le 2^{\ell-1}$.
Furthermore, as $k\in \SUBPATHUPD(W,i,I,i,I,s/2)\cup\{i,I\}$, we get, by
Lemma~\ref{L-SUBPATHUPD}, that $f_{ik}=\del(i,k)$ and
$f_{kI}=\del(k,I)$. (The fact that $f_{ik}=\del(i,k)$ follows also from
the induction hypothesis, as $\eta(i,k)\le 2^{\ell-1}$.)
As $\eta(i,I),\eta(i,k)\le 2^{\ell-1}$, we get that
$|\del(i,I)|,|\del(i,k)|\le 2^{\ell-1}M$. Thus
$|\del(k,I)|=|\del(i,I)-\del(i,k)| \le |\del(i,I)|+|\del(i,k)| \le
2^\ell M$. To sum up, we have
$$\begin{array}{ccccc}
    f_{ik}=\del(i,k) & , & f_{kI}=\del(k,I) & , & f_{Ij}=\del(I,j) \\
    |f_{ik}|\le 2^{\ell-1}M & , & |f_{kI}|\le 2^{\ell}M & , &
    |f_{Ij}|\le 2^{\ell-1}M \\
\end{array}$$
As $k\in B$ and $I,j\in V$, 
after the first distance product of the $\ell$-th iteration, we get
that
$$f_{kj}\le f_{kI}+f_{Ij}=\del(k,I)+\del(I,j)=\del(k,j)\;,$$
and thus
$f_{kj}=\del(k,j)$ and $|f_{kj}|<2^{\ell+1}M$.  As $i,j\in V$ and $k\in
B$, after the second distance product we get that
$$f_{ij}\le f_{ik}+f_{kj}=\del(i,k)+\del(k,j)=\del(i,j)\;,$$
and thus $f_{ij}=\del(i,j)$, as required.
\end{proof}

Finally, we describe the changes that should be made to \SHORTPATH\ if
we want it to detect negative weight cycles, and continue to work in the
presence of zero weight cycles. Detecting negative weight cycles is
easy. We simply check, after each iteration, whether $f_{ii}<0$, for
some $i\in V$. Making \SHORTPATH\ work in the presence of zero weight
cycles requires more substantial changes. 

Before describing the changes required, let us review the problems
caused by zero weight cycles. First, as mentioned in
Section~\ref{S-paths}, the shortest paths returned by $\PATH(W,i,j)$ are
not necessarily simple. Thus, calls to $\SUBPATH(W,i,j,s)$
and $\SUBPATHUPD(W,i,j,i,j,s)$ may return multisets with less than~$s$
distinct elements. As a consequence, the bridging set returned by
$\FINDBRIDGE(W,s)$ and by $\FINDBRIDGEUPD(F,W,s)$ are not necessarily of
size $O(n\log n/s)$. Second, Lemma~\ref{L-close}, that plays a crucial
role in the correctness proof of algorithm \SHORTPATH\ no longer holds
in the presence of zero weight cycles.

To fix these problems we use an approach that is similar to the one used
in Section~\ref{S-paths}. After each iteration of \SHORTPATH\ we call
algorithm \WITTOSUC\ convert the matrix of witnesses~$W$ into a
matrix~$S$ of successors. As the complexity of \WITTOSUC\ is $O(n^2)$,
the extra cost involved is negligible. Even though $W$ does not describe
yet shortest paths between all pairs of vertices of the graph, it is not
difficult to verify that if for some $i,j\in V$ the matrix $W$ describes
a shortest path from~$i$ to~$j$ in the graph, then~$S$ would describe a
{\em simple\/} shortest path from~$i$ to~$j$. Using~$S$ instead of~$W$,
it is then easy to find, in $O(s)$ time, the {\em first\/} $s$
intermediate vertices on a shortest path from~$i$ to~$j$. The bridging
set returned by \FINDBRIDGEUPD\ would then satisfy the condition of
Lemma~\ref{L-close} and the correctness of the algorithm would follow.

\section{Almost shortest paths}
\label{S-approx}

In this section we show that estimations with a relative error of at
most~$\eps$ of all the distances in a weighted directed graph on~$n$
vertices with {\em non-negative\/} integer weights bounded by~$M$ can
be computed deterministically in $\Ot((n^\omega/\eps)\cdot\log M)$ time. If
the weights of the graphs are non-integral, we can scale them so that
the minimal non-zero weight would be 1, multiply them by $1/\eps$,
round them up and then run algorithm with the integral weights
obtained. The running time of the algorithm would then be
$\Ot((n^\omega/\eps)\cdot\log (W/\eps))$, as claimed in the abstract and
in the introduction.

For unweighted directed graphs, it is easy to obtain such estimates
in $\Ot(n^\omega/\eps)$ time. Let~$A$ be the adjacency matrix of the
graph and let $\eps>0$. By computing the Boolean matrices 
$A^{\lfloor(1+\eps)^\ell \rfloor}$ and
$A^{\lceil(1+\eps)^\ell \rceil}$, 
for every $0\le\ell\le \log_{1+\eps}n$, we can easily obtain estimates with
a relative error of at most~$\eps$.
The time required to compute all these matrices is $\Ot(n^\omega/\eps)$.
We next show that almost the same time bound can be obtained when the
graph is weighted. The algorithm is again quite simple.

The main idea used to obtain almost shortest paths is {\em scaling}. A
very simple scaling algorithm, called \SCALE, is given in Figure~\ref{F-scale}.
The algorithm receives an $n\times n$ matrix~$A$ containing {\em
   non-negative\/} elements. It returns an $n\times n$ matrix~$A'$.
The elements of~$A$ that lie in the interval $[0,M]$ are
scaled, and rounded up, into the $R+1$ different values
$0,1,\ldots,R$. We refer to~$R$ as the {\em resolution\/} of the scaling.

\begin{figure}[t]
\begin{center}
\framebox{\hspace{0.6cm}\parbox{4.8in}{
{\tt algorithm} $\SCALE(A,M,R)\phantom{2^{2^{2^2}}}$ \\[3pt]
$a'_{ij}\gets
\cases{\lceil Ra_{ij}/M \rceil & if $\;0\le a_{ij}\le M$ \cr
+\infty & otherwise\cr}$ \\[3pt]
Return $A'$. $\phantom{2_{2_{2_2}}}$
}\hspace{0.6cm}}
\end{center}

\caption{\label{F-scale}A simple scaling algorithm.}
\end{figure}

We next describe a simple algorithm for computing {\em approximate\/}
distance products. The algorithm, called \APPROXDISTPROD, is given in
Figure~\ref{F-approx-dist-prod}. It receives two
matrices~$A$ and~$B$ whose elements are non-negative integers.
It uses {\em adaptive scaling\/} to compute a very accurate
approximation of the distance product of 
$A$ and~$B$.

\begin{figure}[t]

\newcommand{\RR}{R}
\newcommand{\rr}{r}
\newcommand{\MM}{M}

\begin{center}
\framebox{\hspace{0.4cm}\parbox{4.8in}{
\algorithm\ 
\makebox[0pt][l]{$\vphantom{2^{2^{2^2}}}\APPROXDISTPROD(A,B,M,R)$} \\[3pt]
$C\gets +\infty$ \\[3pt]
{\tt for} $r\gets \lfloor\log_2 R\rfloor$ {\tt to} 
$\lceil\log_2 M\rceil$ {\tt do}\\
{\tt begin}\\[3pt]
$\null\qquad A' \gets \SCALE(A,2^r,R)$\\
$\null\qquad B' \gets \SCALE(B,2^r,R)$\\
$\null\qquad C' \gets \DISTPROD(A',B',R)$\\
$\null\qquad C \gets \min\{\;C\;,\;(2^r/R)\cdot C'\;\}$\\
{\tt end}\\[3pt]
{\tt return} $C$ $\phantom{2_{2_{2_2}}}$
}\hspace{0.6cm}}
\end{center}
\caption{\label{F-approx-dist-prod} Approximate distance products.}
\end{figure}

\newcommand{\bc}{\bar{c}}

\begin{lemma}\label{L-approx-dist-prod}
    Let $\bar{C}$ be the distance product of the matrices obtained from
    the matrices~$A$ and~$B$ by replacing the elements that are larger
    than~$M$ by~$+\infty$.
    Let~$M$ and~$R$ be powers of two. Let $C$ be the matrix
    obtained by calling $\APPROXDISTPROD(A,B,M,R)$. Then, for every
    $i,j$ we have $\bc_{ij}\le c_{ij} \le (1+\frac{4}{R})\bc_{ij}$.
\end{lemma}

\begin{proof}
    The inequalities $\bc_{ij}\le c_{ij}$ follow from the fact that
    elements are always rounded upwards by \SCALE. We next show that
    $c_{ij}\le (1+\frac{4}{R})\bc_{ij}$. Let $k$ be a witness for
    $\bc_{ij}$, i.e., $\bc_{ij}=a_{ik}+b_{kj}$. Assume,
    without loss of generality, that $a_{ik}\le b_{kj}$.
    Suppose that $2^{s-1}<
    b_{kj}\le 2^{s}$, where $1\le s\le \log_2 M$ (the cases $b_{kj}=0$
    and $b_{kj}=1$ are easily dealt with separately). If
    $s\le \log_2 R$, then in the first iteration of
    \APPROXDISTPROD, when $r=\log_2 R$, we get $c_{ij}=\bc_{ij}$. Assume,
    therefore, that $\log_2 R\le s\le \log_2 M$.
    In the iteration of \APPROXDISTPROD\ in
    which $r=s$ we get that 
    $$ \frac{2^r\ct a'_{ik}}{R}\le a_{ik} + \frac{2^r}{R}\quad,\quad 
       \frac{2^r\ct b'_{kj}}{R}\le b_{kj} + \frac{2^r}{R}\;.$$
    Thus, after the call to \DISTPROD\ we have
    $$ c_{ij} \le \frac{2^r\ct a'_{ik}}{R} + \frac{2^r\ct b'_{jk}}{R} \le 
       a_{ik} + b_{kj} + \frac{2^{r{+}1}}{R} \le (1+\frac{4}{R})\bc_{ij}\;,$$
    as required.
\end{proof}

If~$A$ and~$B$ are two $n\times n$ matrices, then the complexity of
\APPROXDISTPROD\ is $\Ot(R\ct n^\omega\ct\log M)$. As we will usually
have $R\ll M$, algorithm \APPROXDISTPROD\ will usually be much faster
than \DISTPROD, whose complexity is $\Ot(M\ct n^\omega)$.

Algorithm \APPROXSHORTPATH, given in Figure~\ref{F-approx}, receives as
input an $n\times n$ matrix~$D$ representing the non-negative edge
weights of a directed graph on~$n$ vertices, and an error bound $\eps$.
It computes estimates, with a stretch of at most $1+\eps$, of all
distances in the graph. Algorithm \APPROXSHORTPATH\ starts by letting
$F\gets D$. It then simply squares~$F$, using distance products,
$\lceil\log_2 n\rceil$ times. Rather than compute these distance
products exactly, it uses \APPROXDISTPROD\ to obtain very accurate
approximations of them.

Algorithm \APPROXSHORTPATH\ uses a resolution~$R$ which is the smallest
power of two greater than or equal to $4\lceil\log_2
n \rceil/\ln(1+\eps)$. Thus, $R=O((\log n)/\eps)$.
Using Lemma~\ref{L-approx-dist-prod}, it is easy to
show by induction that the stretch of the elements of~$F$ after the
$\ell$-th iteration is at most $(1+\frac{4}{R})^\ell$. After $\lceil
\log_2 n\rceil$ iterations, the stretch of the elements of $F$ is at
most 
$$ \left(1+\frac{4}{R}\right)^{\lceil\log_2 n\rceil} \le 
\left(1+\frac{\ln(1+\eps)}{\lceil\log_2 n\rceil}\right)^{\lceil\log_2
   n\rceil}  \le 
1+\eps\;.$$ 

As $R=O((\log n)/\eps)$, the complexity of each approximate distance
product computed by \APPROXSHORTPATH\ is $\Ot((n^\omega/\eps)\ct\log M)$. As
only $\lceil\log_2 n\rceil$ such products are computed, this is also the
complexity of the whole algorithm. We have thus established:

\begin{figure}[t]

\begin{center}
\framebox{\hspace{0.6cm}\parbox{4.8in}{
\algorithm\ $\mbox{\APPROXSHORTPATH}(D,\eps) \phantom{2^{2^{2^2}}}$ \\[3pt]
$F\gets D$ \\
$M\gets \max\{\;d_{ij} : d_{ij}\ne+\infty\}$\\
$R\gets 4\lceil\log_2n\rceil/\ln(1+\eps)$ \\ 
$R\gets 2^{\lceil\log_2 R\rceil}$\\[3pt]
{\tt for} $\ell\gets 1$ {\tt to} $\lceil\log_2 n\rceil$ {\tt do}\\
{\tt begin}\\[3pt]
$\null\qquad F' \gets \mbox{\APPROXDISTPROD}(F,F,Mn,R)$\\
$\null\qquad F \gets \min\{\;F\;,\;F'\;\}$\\
{\tt end}\\[3pt]
{\tt return} $F$ $\phantom{2_{2_{2_2}}}$
}\hspace{0.6cm}}
\end{center}

\caption{\label{F-approx} Approximate shortest paths.}
\end{figure}

\begin{theorem}\label{T-approx}
    Algorithm \APPROXSHORTPATH\ runs in $\Ot((n^\omega/\eps)\ct\log M)$
    time and produces a matrix of estimated distances with a relative
    error of at most $\eps$.
\end{theorem}

As described, algorithm \APPROXSHORTPATH\ finds approximate distances. 
It is easy to modify it so that it would also return a matrix~$W$ of
witnesses using which approximate shortest paths could also be found.

\section{Concluding remarks}
\label{S-concl}

\ignore{
The algorithm we presented for solving the APSP problem for
directed graphs with small integer weights is randomized. We
believe that the algorithm can be derandomized, with only a
small loss of efficiency. We have made some progress in
this direction.
}

The results of Seidel \cite{Seidel95} and 
Galil and Margalit \cite{GaMa97a},\cite{GaMa97b} show that
the complexity of the APSP problem for unweighted {\em
undirected\/} graphs is $\Ot(n^\omega)$. The exact
complexity of the directed version of the problem is not
known yet. In view of the results contained in this paper,
there seem to be two plausible conjectures. The first is
$\Ot(n^{2.5})$. The second is $\Ot(n^\omega)$. Galil and
Margalit \cite{GaMa97a} conjecture that the problem for
directed graphs is {\em harder\/} than the problem for
undirected graphs. Proving, or disproving, this conjecture
is a major open problem.

Another interesting open problem is finding the maximal
value of~$M$ for which the APSP problem with integer weights
of absolute value at most~$M$ can be solved in sub-cubic time.
Our algorithm runs in sub-cubic time for $M<n^{3-\omega}$,
as does the algorithm of Takaoka \cite{Takaoka98}. Can the
APSP problem be solved in sub-cubic time, for example, when $M=n$?

Finally, we note that the shortest paths returned by the algorithms
presented in this paper do not necessarily use a minimum number of
edges. Producing shortest paths that do use a minimum number of edges
seems to be a slightly harder problem. For more details, see Zwick
\cite{Zw99ST}.

\ignore{
We also presented an algorithm for obtaining almost exact
solutions to the APSP problem for directed graphs with
arbitrary non-negative real weights. For every desired
accuracy $\eps$, the algorithm performs only a
polylogarithmic number of multiplications of matrices whose
elements are small integers. 
}

\section*{Acknowledgment}
I would like to thank Victor Pan for sending me a preprint of
\cite{HuPa98} and for answering several questions regarding matrix
multiplication.


\begin{thebibliography}{{Chv}79}

\bibitem[ACIM99]{AiChInMo99}
D.~Aingworth, C.~Chekuri, P.~Indyk, and R.~Motwani.
\newblock Fast estimation of diameter and shortest paths (without matrix
  multiplication).
\newblock {\em SIAM Journal on Computing}, 28:1167--1181, 1999.

\bibitem[AGM97]{AlGaMa97}
N.~Alon, Z.~Galil, and O.~Margalit.
\newblock On the exponent of the all pairs shortest path problem.
\newblock {\em Journal of Computer and System Sciences}, 54:255--262, 1997.

\bibitem[AHU74]{AHU74}
A.V. Aho, J.E. Hopcroft, and J.D. Ullman.
\newblock {\em The design and analysis of computer algorithms}.
\newblock Addison-Wesley, 1974.

\bibitem[AN96]{AlNa96}
N.~Alon and M.~Naor.
\newblock Derandomization, witnesses for {Boolean} matrix multiplication and
  construction of perfect hash functions.
\newblock {\em Algorithmica}, 16:434--449, 1996.

\bibitem[BCS97]{BuClSh97}
P.~Burgisser, M.~Clausen, and M.A. Shokrollahi.
\newblock {\em Algebraic complexity theory}.
\newblock Springer-Verlag, 1997.

\bibitem[{Chv}79]{Chvatal79}
V.~{Chv\'{a}tal}.
\newblock A greedy heuristic for the set-covering problem.
\newblock {\em Mathematics of Operations Research}, 4:233--235, 1979.

\bibitem[CLR90]{CoLeRi90}
T.H. Cormen, C.E. Leiserson, and R.L. Rivest.
\newblock {\em Introduction to algorithms}.
\newblock The MIT Press, 1990.

\bibitem[Cop97]{Coppersmith97}
D.~Coppersmith.
\newblock Rectangular matrix multiplication revisited.
\newblock {\em Journal of Complexity}, 13:42--49, 1997.

\bibitem[CW90]{CopWin90}
D.~Coppersmith and S.~Winograd.
\newblock Matrix multiplication via arithmetic progressions.
\newblock {\em Journal of Symbolic Computation}, 9:251--280, 1990.

\bibitem[CZ97]{CoZw97}
E.~Cohen and U.~Zwick.
\newblock All-pairs small-stretch paths.
\newblock In {\em Proceedings of the 8th Annual ACM-SIAM Symposium on Discrete
  Algorithms, New Orleans, Louisiana}, pages 93--102, 1997.

\bibitem[DHZ00]{DoHaZw00}
D.~Dor, S.~Halperin, and U.~Zwick.
\newblock All pairs almost shortest paths.
\newblock {\em SIAM Journal on Computing}, 29:1740--1759, 2000.

\bibitem[Dij59]{Di59}
E.W. Dijkstra.
\newblock A note on two problems in connexion with graphs.
\newblock {\em Numerische Mathematik}, 1:269--271, 1959.

\bibitem[Fre76]{Fredman76}
M.L. Fredman.
\newblock New bounds on the complexity of the shortest path problem.
\newblock {\em SIAM Journal on Computing}, 5:49--60, 1976.

\bibitem[FT87]{FrTa87}
M.L. Fredman and R.E. Tarjan.
\newblock Fibonacci heaps and their uses in improved network optimization
  algorithms.
\newblock {\em Journal of the ACM}, 34:596--615, 1987.

\bibitem[GM93]{GaMa93}
Z.~Galil and O.~Margalit.
\newblock Witnesses for boolean matrix multiplication.
\newblock {\em Journal of Complexity}, 9:201--221, 1993.

\bibitem[GM97a]{GaMa97a}
Z.~Galil and O.~Margalit.
\newblock All pairs shortest distances for graphs with small integer length
  edges.
\newblock {\em Information and Computation}, 134:103--139, 1997.

\bibitem[GM97b]{GaMa97b}
Z.~Galil and O.~Margalit.
\newblock All pairs shortest paths for graphs with small integer length edges.
\newblock {\em Journal of Computer and System Sciences}, 54:243--254, 1997.

\bibitem[HP98]{HuPa98}
X.~Huang and V.Y. Pan.
\newblock Fast rectangular matrix multiplications and applications.
\newblock {\em Journal of Complexity}, 14:257--299, 1998.

\bibitem[Joh77]{Jo77}
D.B. Johnson.
\newblock Efficient algorithms for shortest paths in sparse graphs.
\newblock {\em Journal of the ACM}, 24:1--13, 1977.

\bibitem[KKP93]{KaKoPh93}
D.R. Karger, D.~Koller, and S.J. Phillips.
\newblock Finding the hidden path: time bounds for all-pairs shortest paths.
\newblock {\em SIAM Journal on Computing}, 22:1199--1217, 1993.

\bibitem[Lov75]{Lovasz75}
L.~Lov{\'{a}}sz.
\newblock On the ratio of optimal integral and fractional covers.
\newblock {\em Discrete Mathematics}, 13:383--390, 1975.

\bibitem[McG95]{McGeoch95}
C.C. McGeoch.
\newblock All-pairs shortest paths and the essential subgraph.
\newblock {\em Algorithmica}, 13:426--461, 1995.

\bibitem[Pan85]{Pan85}
V.~Pan.
\newblock {\em How to multiply matrices faster}.
\newblock Lecture notes in computer science, volume 179. Springer-Verlag, 1985.

\bibitem[Sei95]{Seidel95}
R.~Seidel.
\newblock On the all-pairs-shortest-path problem in unweighted undirected
  graphs.
\newblock {\em Journal of Computer and System Sciences}, 51:400--403, 1995.

\bibitem[SS71]{SchSt71}
A.~{Sch\"{o}nhage} and V.~Strassen.
\newblock Schnelle multiplikation grosser zahlen.
\newblock {\em Computing}, 7:281--292, 1971.

\bibitem[SZ99]{ShZw99}
A.~Shoshan and U.~Zwick.
\newblock All pairs shortest paths in undirected graphs with integer weights.
\newblock In {\em Proceedings of the 40th Annual IEEE Symposium on Foundations
  of Computer Science, New York, New York}, pages 605--614, 1999.

\bibitem[Tak92]{Takaoka92}
T.~Takaoka.
\newblock A new upper bound on the complexity of the all pairs shortest path
  problem.
\newblock {\em Information Processing Letters}, 43:195--199, 1992.

\bibitem[Tak98]{Takaoka98}
T.~Takaoka.
\newblock Subcubic cost algorithms for the all pairs shortest path problem.
\newblock {\em Algorithmica}, 20:309--318, 1998.

\bibitem[Tho99]{Thorup99}
M.~Thorup.
\newblock Undirected single-source shortest paths with positive integer weights
  in linear time.
\newblock {\em Journal of the ACM}, 46:362--394, 1999.

\bibitem[Tho00]{Thorup00}
M.~Thorup.
\newblock Floats, integers, and single source shortest paths.
\newblock {\em Journal of Algorithms}, 35:189--201, 2000.

\bibitem[Yuv76]{Yuval76}
G.~Yuval.
\newblock An algorithm for finding all shortest paths using ${N}^{2.81}$
  infinite-precision multiplications.
\newblock {\em Information Processing Letters}, 4:155--156, 1976.

\bibitem[Zwi98]{Zw98FO}
U.~Zwick.
\newblock All pairs shortest paths in weighted directed graphs -- exact and
  almost exact algorithms.
\newblock In {\em Proceedings of the 39th Annual IEEE Symposium on Foundations
  of Computer Science, Palo Alto, California}, pages 310--319, 1998.

\bibitem[Zwi99]{Zw99ST}
U.~Zwick.
\newblock All pairs lightest shortest paths.
\newblock In {\em Proceedings of the 31th Annual ACM Symposium on Theory of
  Computing, Atlanta, Georgia}, pages 61--69, 1999.

\end{thebibliography}

\end{document}